\documentclass[aps,pra,twocolumn]{revtex4-1}
\usepackage{graphicx}

\usepackage{epstopdf}

\usepackage{bm}
\usepackage{dcolumn}
\usepackage{latexsym} 

\newcommand{\D}{\,\mathrm{d}}
\begin{document}


\title{
  Quantum gravity slingshot: orbital precession due
  to the modified uncertainty principle, from analogs to tests of Planckian physics with quantum fluids}
\author{Giulia Marcucci}
\affiliation{Department of Physics, University Sapienza, Piazzale Aldo Moro 2, 00185, Rome (IT)}
\author{Claudio Conti}
\affiliation{Department of Physics, University Sapienza, Piazzale Aldo Moro 2, 00185, Rome (IT)}
\email{claudio.conti@uniroma1.it}
\date{\today}

\begin{abstract}
Modified uncertainty principle and non-commutative variables may phenomenologically account for quantum gravity effects, independently of the considered theory of quantum gravity. 
We show that quantum fluids enable experimental analogs and direct tests of the modified uncertainty principle expected to be valid at the Planck scale.
We consider a quantum clock realized by a long-lasting quantum fluid wave-packet orbiting in a trapping potential.
We investigate the hydrodynamics of the Schr\"odinger equation encompassing kinetic terms due to Planck-scale effects.
We study the resulting generalized mechanics and validate the predictions by quantum simulations.
Wave-packet orbiting generates a continuous amplification of the quantum gravity effects. The non-commutative variables in the phase-space produce a precession and an acceleration of the orbital motion. 
The precession of the orbit is strongly resembling the famous orbital precession of the perihelion of Mercury used by Einstein to validate the corrections of general relativity to Newton's theory. In our case, the corrections are due to the modified uncertainty principle.
The results can be employed to emulate quantum gravity in the laboratory, or to realize human-scale experiments to determine bounds for the most studied quantum gravity models and probe Planckian physics.
\end{abstract}


\maketitle
\section{Introduction}
The phenomenology of the modified fundamental laws of physics at the Planck scale attracts a large community of scientists.~\cite{Hossenfelder2013,  AMELINOCAMELIA2001255, Gambini99, MukhanovBook2007, BookRovelli2015, ThiemannBook2007, BookKiefer2012,Fedi2018} The challenge is to identify feasible experiments to test the apparently inaccessible limits of quantum mechanics and general relativity at the Planck scale. For example, very recently the possible observation of Planckian corrections to general relativity by space-based interferometers was foreseen.~\cite{PhysRevLett.120.081101}
The difficulties in realizing such large-scale experiments trigger studying quantum simulations. The simulations are realizable in Earth-based laboratories in a human-life timescale. 
Beyond mimicking quantum gravity theories, researchers also aim at realizing analog experimental models of black holes~\cite{PhysRevLett.85.4643,PhysRevA.80.065802,Faccio2017}, Hawking radiation~\cite{Steinhauer2016,faccio2012,unruh1976,PhysRevA.97.013823}, inflation and universe expansion~\cite{Fischer2017, Eckel2018}, dark-matter models~\cite{Paredes2016},  and related phenomena.~\cite{Zheng06,Barcelo2003,Longhi2011}
The analogy is a fundamental tool in physics, and experimental and theoretical analogs may deepen our understanding of quantum gravity theories,~\cite{Braidotti2017} and of other challenging proposals as time-asymmetric quantum mechanics~\cite{Gentilini2015Glauber, PhysRevA.94.052136,PhysRevA.60.861}.

A large amount of literature deals with the tantalizing need to modify the uncertainty principle and related non-commutative geometry.~\cite{Kempf95}
In standard quantum mechanics, there is no minimal value for the position uncertainty $\Delta X$. However, theories in quantum gravity imply the existence of a minimal length scale, commonly (but not necessarily) identified with the Planck length $\ell_P$.~\cite{Hossenfelder2013} Hence a lower bound for $\Delta X$ must be included in quantum mechanics, and the Heisenberg relation $\Delta X\Delta P\geq \hbar/2 $ with the momentum uncertainty $\Delta P$ must be generalized - also in the non-relativistic limit considered hereafter.
The most accepted formulation of the so-called modified or generalized uncertainty principle (GUP) reads~\cite{Garay1995, KempfMangano97}
\begin{equation}
  \Delta X\Delta P>\frac{\hbar}{2}\left(1+\beta \Delta P^2\right).
  \label{GUP}
  \end{equation}
Eq.~(\ref{GUP}) implies $\Delta X>\hbar\sqrt{\beta}$ with $\beta$ a unknown constant, which is eventually related to $\ell_P$. We need experimental evidence to assess the validity of Eq.~(\ref{GUP}) and fix bounds for the value of $\beta$.

The generalized uncertainty principle in Eq.~(\ref{GUP}) arises in a theory-independent-way, that is, many theories attempting to unify gravity and quantum mechanics - including string theory~\cite{Witten1996} - predict modifications of the Heisenberg principle as in Eq.~(\ref{GUP}). The generalized uncertainty principle is also related to modified commutation relations and to non-commutative theories.~\cite{Connes,Maggiore1993,ACamelia13,Douglas2001,Ghosh2014}

In a general perspective, the challenge is to modify non-relativistic quantum mechanics to account for Eq.~(\ref{GUP}), and study the resulting ``GUP phenomenology'': new physical effects to experimentally confirm the validity of the modified Heisenberg principle and of the quantum theories of gravity.
The simplest approach - without resorting to the details of quantum gravity theories - is to introduce corrections to quantum mechanics in order to account for the minimal-length scenario. This methodology aims at predicting and quantifying observable phenomena related to the generalized uncertainty principle, and this strategy is justified by the fact that any quantum gravity theory in the low-energy limit will result in a modified quantum mechanics.

A generalized uncertainty principle has many phenomenological implications as, for example, in cosmological dynamics, black-body radiation, wave-packet localization and related investigations reported by several authors.~\cite{Scardigli99,Chang02, Das08,Pedram12, Ghosh2014}
Quantitative bounds for the $\beta$ parameter in the modified Heisenberg principle in Eq.~({\ref{GUP}}) were also reported.~\cite{Das08}
Other authors discussed human-scale laboratory tests with optomechanical and orbiting classical objects~\cite{Pikovski12, Bekenstein12, Marin2015, Silagadze2009}. However, the application of generalized quantum mechanics to the macroscopic world is questionable.~\cite{ACamelia13}
Analogs, i.e., physical systems governed by laws mathematically identical to those of the generalized quantum mechanics, were considered in the fields of optics and relativistic Bose-Einstein condensates.~\cite{Pedram2011,Conti2014,Castellanos2017,Braidotti201734, Faizal2018}

\begin{figure}
\includegraphics[width=\columnwidth]{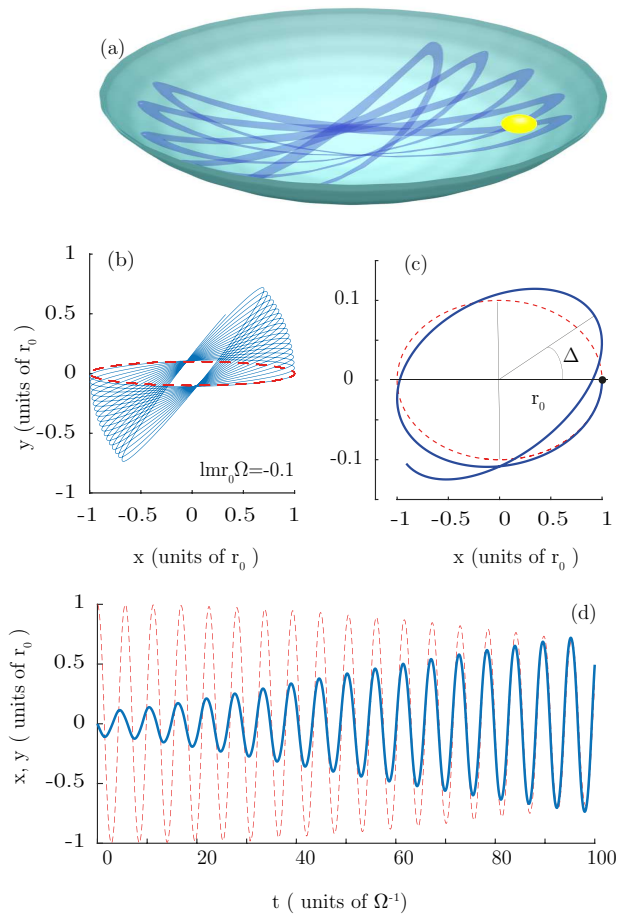}
\caption{
(Color online)  (a) A quantum-fluid wave-packet (yellow) in an elliptical orbit in a parabolic potential; the quantum gravity perturbations to the standard non-relativistic quantum mechanics induce a orbital precession;   (b) solutions of Eqs.~(\ref{odebeta}) with
$l m r_0\Omega=-0.1$, $\beta (m r_0 \Omega)^2=0.1$ (continuous line) and $\beta=0$ (dashed line) with $\Omega t$ in the range $[0,100]$;
(c) as in (a) for $\Omega t\leq 8$, the dot indicates the initial position;
(d) $x$ (dashed line) and $y$ versus time $t$.
\label{figure_prec}}
\end{figure}

In this manuscript, we study the orbital precession of a quantum fluid (Figure~\ref{figure_prec}a) due to the perturbation to quantum mechanics induced by quantum gravity and - more specifically - due to the additional kinetic terms that account for the generalized uncertainty principle as in Eq.~(\ref{GUP}).  Quantum fluids are studied in the vast literature concerning Bose-Einstein condensates (BECs), quantum nonlinear optics and polaritonics.~\cite{Carusotto2013, Klaers2010, PhysRevA.90.043853, Dominici2015, Amo2009}
We consider a quantum fluid wave-packet with a non-vanishing angular momentum in a trapping potential. We show that the wave-packet elliptical orbit is perturbed by the quantum gravity effects. The longtime observation of many orbits evidences a precession and a delay.
During orbiting, GUP phenomenology is amplified and this amplification (``quantum gravity slingshot'') may allow laboratory analogs and real experimental tests of Planckian physics with quantum fluids.
\section{Generalized Schr\"odinger equation}
\noindent We study the generalized Schr\"odinger - or Gross-Pitaevskii - equation (SE) for quantum fluids with trapping potential $V$ in two or three spatial dimensions
\begin{equation}
\imath \hbar \frac{\partial \psi}{\partial t}=-\frac{\hbar^2}{2m}\nabla^2 \psi+\frac{\beta \hbar^4}{m}\nabla^4 \psi + V(\mathbf{r}) \psi=\hat H \psi\text{.}
\label{eq1}
\end{equation}
Eq.~(\ref{eq1}) describes a non-interacting atomic BEC, realized by employing the Feshbach resonance.~\cite{RevModPhys.82.1225} Eq.~(\ref{eq1}) also describes a polaritonic or photonic condensate. In the latter case, as extensively discussed in the literature, $t$ corresponds to the propagation direction, and $\hbar$ must be replaced by the reduced light wavelength $\lambda/2\pi$.~\cite{Longhi2018}
Eq.~(\ref{eq1}) contains a kinetic term weighted by $\beta$.
As considered by various authors~\cite{Das08,Das09,Conti2014}, the additional kinetic term is the simplest modification to the Schr\"odinger equation that implies the generalized uncertainty principle in Eq.~(\ref{GUP}) and arises from fundamental modifications to the geometry of the space-time at the Planck scale, which
change the dispersion relation of free particles, as photons. In the optical case, corrections to paraxial approximation introduce the $\nabla^4 \psi$ term, and the ratio between $\lambda$ and the beam waist determines $\beta$ in a optical analogue to the generalized quantum mechanics.~\cite{Conti2014} For cold-atoms, higher order derivatives may arise from relativistic effects (not considered here), when the fluid velocity is comparable with the velocity of light $c$.~\cite{Briscese12} Another mechanism is the modification of the dispersion relation in models as doubly-special-relativity.~\cite{Mercati10}
\begin{figure}
\includegraphics[width=\columnwidth]{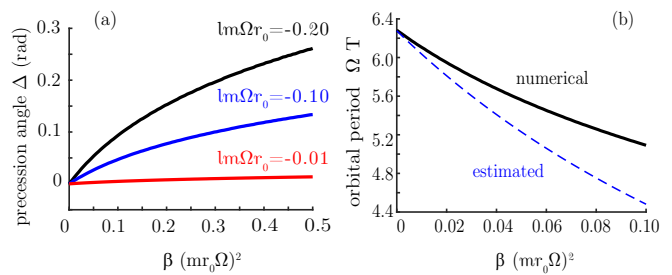}
\caption{ (Color online) Effect of the perturbation $\beta$ to the uncertainty principle in Eq.~(\ref{GUP}) on the (a) precession angle $\Delta$ for various $l$ and (b)   orbital period $T$ for $lm\Omega r_0=1.01$, the dashed line is the  theoretical estimate (see text).
\label{figure_orbit}}
\end{figure}

Letting  $\hat p=-\imath \hbar \nabla$, the Hamiltonian in Eq.~(\ref{eq1}) is
\begin{equation}
\hat H \psi=\frac{\hat p^2}{2m}\psi+\frac{\beta}{m}\hat p^4 \psi +V(\mathbf r) \psi
\end{equation}
with the position vector $\mathbf{r}=\left(x,y,z\right)=\left(x^1,x^2,x^3\right)$. One introduces non-commutative coordinates by a new set of ``high-energy'' variables $(X^{\mu},P^{\mu})$, which, in the simplest case considered here, read ($\mu,\nu=1,2,3$)
\begin{equation}
\begin{array}{l}
\hat X^\mu=\hat x^\mu;\\
\hat P^\nu=\hat p^\nu(1+\beta\hat p^2)\text{.}
\end{array}
\label{NC1}
\end{equation}
By $[\hat x^\mu,\hat x^\nu]=0$, $[\hat p^\mu,\hat p^\nu]=0$, $[\hat x^\mu,\hat p^\nu]=\imath\hbar\delta^{\mu,\nu}$, one has $Â£\left[\hat P^\mu,\hat P^\nu\right]=\left[\hat X^\mu,\hat X^\nu\right]=0$ and~\cite{PhysRevD.66.026003,Das08, Silagadze2009}
\begin{equation}
\begin{array}{l}
\left[\hat X^\mu,\hat P^{\nu}\right]=\imath \hbar(\delta^{\mu \nu}+\beta \delta^{\mu \nu} \hat P^2+2\beta \hat P^{\mu} \hat P^{\nu})\text{.}\\
\end{array}
\label{noncomm1}
\end{equation}
This is an example of non-commutative phase-space coordinates: at variance with standard quantum mechanics, the momentum
and the position in different directions do not commute. By $\hat P^{\mu}$, Eq.~(\ref{eq1}) has the traditional form 
\begin{equation}
\imath \hbar \psi_t=\frac{\hat P^2}{2 m}\psi +V(\mathbf r) \psi,
\end{equation}
but the commutation relations are modified and Eq.~(\ref{GUP}) holds true.
The simplest effect of the additional kinetic term are shifts $\Delta E_k$ in the energetic levels
of the eigenstates,~\cite{Brau1999} $\Delta E_k=\frac{\beta}{m}\langle k|\hat p^4| k \rangle=4 m \beta \langle k| V^2 | k \rangle$.
Such perturbations may eventually occur in optical and atomic quantum fluids, however they are very difficult to observe.~\cite{Das08} We consider a more accessible phenomenology related to the dynamics of a wave-packet orbiting in the potential.
\section{Hydrodynamic limit and Hamilton equations}
We study a wave-packet initially located at $r=\sqrt{x^2+y^2}=r_0$ that rotates with a non-vaninshing angular momentum (Fig.~\ref{figure_prec}a). The trajectory is found in the hydrodynamic approximation~\cite{Brown1972,VanVleck1928} by letting $\psi=A \exp(\imath \mathcal{S}/\hbar)$ with $\hbar\rightarrow 0$.~\cite{Dalfovo99} The Hamilton-Jacobi equation for $\mathcal{S}$ is
\cite{landau2013QuaMecShoCouThePhy, goldsteinbook}
\begin{equation}
  \frac{\partial\mathcal{S}}{\partial t}+\frac{\nabla \mathcal{S}^2}{2m}+\frac{\beta}{m}(\nabla \mathcal{S}^2)^2+V=0\text{.}
  \label{HJ}
\end{equation}
By the classical Hamiltonian $H(\mathbf{q},\mathbf{p})=p^2/2m+(\beta/m)p^4+V$, we have $\partial_t \mathcal{S}+ H(\mathbf{q},\nabla \mathcal{S})=0$.
Eq.~(\ref{HJ}) is solved by the Hamilton system with $\mathbf p=\nabla \mathcal{S}$ ~\cite{goldsteinbook}
\begin{equation}
\begin{array}{l}
\displaystyle \frac{\D\mathbf{q}}{\D t}=\nabla_p H\text{,}\\[10pt]
\displaystyle\frac{\D \mathbf{p}}{\D t}=-\nabla_q H\text{.}
\end{array}
\label{hamilton}
\end{equation}
We consider here a $z-independent$ radial potential with polar symmetry: $V=V(r)$.
Two-dimensional condensates are routinely considered in the literature.~\cite{Dalfovo99,Amo2009,Klaers2010}
In polar coordinates $(r,\theta,z)$, with conjugate momenta $(p_r,p_\theta, p_z)$, we have
\begin{equation}
H=\frac{1}{2m}\left(p_r^2+\frac{p_\theta^2}{r^2}\right)+\frac{\beta}{m}\left(p_r^2+\frac{p_\theta^2}{r^2}\right)^2 +V(r)\text{.}
\label{eq:H}
\end{equation}
The corresponding Lagrangian does not depend explicitly on $\theta$,
hence the conjugate momentum $p_\theta=l$ is conserved,
and the motion occurs in the $z$ plane, with $p_z=0$.
By the conserved $l$, Eqs.~(\ref{hamilton}) are written as
\begin{equation}
\begin{array}{l}
  \displaystyle\frac{\D \theta}{\D t}=\frac{l}{m r^2}\left[1+4\beta\left(p_r^2+\frac{l^2}{r^2}\right)\right]\text{,}\\[10pt]
  \displaystyle\frac{\D p_\theta}{\D t}=0\text{,}\\[10pt]
  \displaystyle\frac{\D r}{\D t}=\frac{p_r}{m}\left[1+4\beta\left(p_r^2+\frac{l^2}{r^2}\right)\right]\text{,}\\[10pt]
  \displaystyle\frac{\D p_r}{\D t}=-\frac{\partial H}{\partial r}=-V'(r)+\frac{l^2}{mr^3}\left[1+4\beta\left(p_r^2+\frac{l^2}{r^2}\right)\right]\text{,}
\end{array}
\label{odebeta}
\end{equation}
with $V'=\D V/\D r$ and, for a parabolic potential,
\begin{equation}V=\frac{1}{2}m\Omega^2 r^2\end{equation}.
\section{Orbital precession and link with the Einstein solution }
Figure~\ref{figure_prec}b shows the numerical solutions of Eqs.~(\ref{odebeta}) with $r(0)=r_0$ and $p_r(0)=0$.
When $\beta=0$ and $l\neq 0$, the orbit is elliptical
(dashed line in Fig.~\ref{figure_prec}b). When $\beta>0$ - continuous line in Fig.~\ref{figure_prec}a - the orbit exhibits a precession (clockwise for $l<0$, and counter-clockwise for $l>0$).
Fig.~\ref{figure_prec}c shows the precession angle $\Delta$ with the particle at $t=0$, $x=r_0$ and $y=0$.

As the orbit rotates, the maximal position in the $y$ coordinate is amplified as shown in Fig.~\ref{figure_prec}d. This orbital enhancement of the quantum gravity effect resembles the known gravity slingshot assist adopted to alter the speed of a spacecraft in orbital mechanics~\cite{bookspaceflight}: the radial acceleration at any turn emphasizes the phenomenology.

We remark that the precession can be related to non-commutative coordinates in the phase-space that arise because of the quantum gravity terms.
If we introduce the classical counter-part of the generalized momenta (\ref{NC1}),
\begin{equation}
  P_{r,\theta}=p_{r,\theta}(1+\beta p^2)
  \end{equation},
  the Hamiltonian $H$ is written as in the case $\beta=0$:
  \begin{equation}
    H=\frac{1}{2m}\left(P_r^2+\frac{P_\theta^2}{r^2}\right)+V(r)\text{.}
  \end{equation}
  However, while the Poisson brackets 
$\{r,p_\theta\}=\{\theta,p_{r}\}=0$ vanish, ~\cite{goldsteinbook} the corresponding quantities for the generalized momenta are
\begin{equation}\{r,P_\theta\}=r^2\{\theta,P_r\}=2\beta P_r P_\theta\text{.}~\label{nonzeroPoisson} \end{equation} Therefore, one has modified mechanics with non-commuting coordinates. In the following, we show that the precession is of the order of magnitude of the brackets in Eq.~(\ref{nonzeroPoisson}), revealing the link between the non-commutative geometry and GUP phenomenology.

\noindent  Eqs.~(\ref{odebeta}) give the precession angle $\Delta$ by
$\D \theta/\D r=l/[p_r(r) r^2]$ with $p_r(r)$ expressed in terms of the conserved quantities $H=E_0$ and $p_\theta=l$. However, no closed form can be found, and Figure~\ref{figure_orbit} shows numerical results. We obtain estimates by considering a nearly circular orbit with $r\simeq r_0$. If $\beta=0$, then $p_\theta=l=m r^2 \dot \theta$ is constant,
and the period is $T_0=2\pi mr_0^2/l$. For $\beta>0$ and $r\simeq r_0$
\begin{equation}
mr_0^2\dot\theta\simeq l\left[1+4\beta\left(p_r^2+\frac{l^2}{r_0^2}\right)\right]\geq l\left(1+4\beta p_0^2\right),
\label{eq:estimate}
\end{equation}
with $p_0^2=\frac{l^2}{r_0^2}$.
In a time interval $\Delta t$, $\Delta \theta\gtrsim \frac{2\pi \Delta t }{T_0}(1+4\beta p_0^2)$, and, when $\Delta t=T_0$, we have the lower bound
\begin{equation}
\Delta=\Delta\theta-2\pi\simeq 8\pi\beta p^2_{0}.
\label{eq:precession}
\end{equation}
The orbital period $T$ is also altered by GUP.
For a $\Delta\theta=2\pi$, the relative variation
\begin{equation}
  \delta T=T/T_0-1\simeq-4\beta p^2_{0}
  \end{equation} is compared with the numerical calculation in Fig.~\ref{figure_orbit}b.

To further validate these arguments, it is also interesting to make a comparison with the analysis of the precession of the perihelion of Mercury, as originally considered by Einstein.~\cite{Einstein1915,CollectedEinstein,MTWbook} This comparison not only allows to derive an additional estimate of the precession, consistent with Eq.~(\ref{eq:precession}), but shows the way the kinetic term induces an effective force, as general relativity induces a correction to the Newton force.
We consider a nearly circular orbit and write the orbit equation starting from the conservation of energy in Eq.~(\ref{eq:H}). By using $\D r/\D \theta=p_r r^2/l$ from Eqs.~(\ref{odebeta}), and $u=1/r$ we have
\begin{equation}
H=\frac{l^2}{2m}\left(\mu+2\beta l^2 \mu^2\right)+\tilde V(u)\text{,}
\label{eq:H1}
\end{equation}
with $\mu\equiv\left(\D u/\D \theta\right)^2+u^2$ and $\tilde V(u)=V(1/u)$.
By deriving Eq.~(\ref{eq:H1}) one obtains, at the lowest order in $\beta$
\begin{equation}
\frac{\D^2 u}{\D \theta^2}+u=-\frac{m}{l^2}\tilde V'(u)(1-4\beta l^2 \mu)
\label{eq:H2}
\end{equation}
with  $\tilde V'(u)=\D \tilde V/\D u$.
Eq.~(\ref{eq:H2}) for $\beta=0$ furnishes the orbit equation;~\cite{goldsteinbook} for $\beta>0$, an additional contribution to the effective potential perturbs the orbit.
\\Eq.~(\ref{eq:H2}) is written as
\begin{equation}
  \frac{\D^2 u}{\D \theta^2}+u=-\frac{m}{l^2}\tilde V'(u)+\beta \mathcal{F}\left(\frac{\D u}{\D \theta},u\right)  
\label{eq:H3}
\end{equation}
with
\begin{equation}
  \mathcal{F}=4m\mu\tilde V'(u)=4m\left[\left(\frac{\D u}{\D \theta}\right)^2+u^2\right]\tilde V'(u)
\end{equation}
representing the perturbation due to the modified uncertainty principle, which disappears for $\beta=0$.

For a nearly circular orbit $u\simeq (1/r_0)$, and a parabolic potential $\tilde V(u)=m\Omega^2/(2u^2)$, the first term in the right hand side of (\ref{eq:H3}) $-m \tilde V'(u)/l^2=m^2 \Omega^2 / (l^2 u^3)$ is nearly a constant, and plays the role of the gravitational field.

In a perturbative expansion in $\beta$, $\mathcal{F}$ in (\ref{eq:H3}) produces a driving force term, as detailed in the following; such a term is in perfect analogy with the Einstein analysis of the Mercury orbit: We consider the solution for $\beta=0$ representing an elliptical orbit with eccentricity $e$:
\begin{equation}
u_0=\sqrt{\frac{m\Omega}{l}}\left[1+e\,\cos(2\theta)\right]\text{.}
\label{eq:orbit1}
\end{equation}
Eq.~(\ref{eq:orbit1}) is valid for the parabolic potential, similar results are obtained for other potentials as, for example, the Newtonian gravitational potential.
The perihelion corresponds to maximal $u$ for  $\theta_n=n\pi$ with $n=0,1,2,....$.

For $\beta>0$, one adopts a perturbative expansion at the lowest order in the eccentricity $e$ and $\beta$.
\\The perturbation force $\mathcal{F}$ reads
\begin{equation}
  \mathcal F=\mathcal F\left(\frac{\D u_0}{\D \theta},u_0\right)\simeq -4 m^{3/2} \Omega^{3/2} l^{1/2}\left[1- e \cos(2 \theta)\right].
\end{equation}
The perturbed orbit Eq.~(\ref{eq:H3}) is
\begin{equation}
\frac{\D u^2}{\D u^2}+u=\frac{m^2\Omega^2}{l^2 u^3}+4\beta \Omega^{3/2}m^{3/2}l^{1/2}\left[1- e \cos(2 \theta)\right]
\label{eq:orbit2}
\end{equation}
The forcing term in (\ref{eq:orbit2}) is directly corresponding to the term obtained by Einstein representing the correction to the orbit due to general relativity.
One can solve Eq.~(\ref{eq:orbit2}) at the lowest order in $\beta$ and $e$ as
\begin{equation}
\displaystyle u=\sqrt{\frac{m\Omega}{l}}\left\{1-\beta m\Omega l+e\left[\cos(2\theta)+\beta \Omega m l \theta\sin(2\theta)\right]\right\}
\label{eq:orbit3}
\end{equation}
By writing Eq.~(\ref{eq:orbit3}) as
\begin{equation}
\displaystyle u\simeq\sqrt{\frac{m\Omega}{l}}\left\{1-\beta m\Omega l+
e \cos\left[2\theta\left(1-\frac{\beta m \Omega l}{2}\right)\right]\right\}\text{,}
\label{eq:orbit3b}
\end{equation}
one sees that the maximum of $u$ in Eq.~(\ref{eq:orbit3b}) occurs for 
\begin{equation}
\theta_n=n\pi\left(1+\frac{\beta m\Omega l}{2}\right)\text{.}
\label{eq:orbit4}
\end{equation}
Hence the perihelion shifts for each half-orbit by an amount
$\pi \beta m\Omega l/2$. Being $l\cong m\Omega r_0^2$ for nearly circular orbits, this result is consistent with the estimate above within numerical factors due to the definition of the precession angle. Eq.~(\ref{eq:orbit4}) shows that the quantum gravity effect accumulates as the precession angle grows with $n$.
\begin{figure*}
\includegraphics[width=1.6\columnwidth]{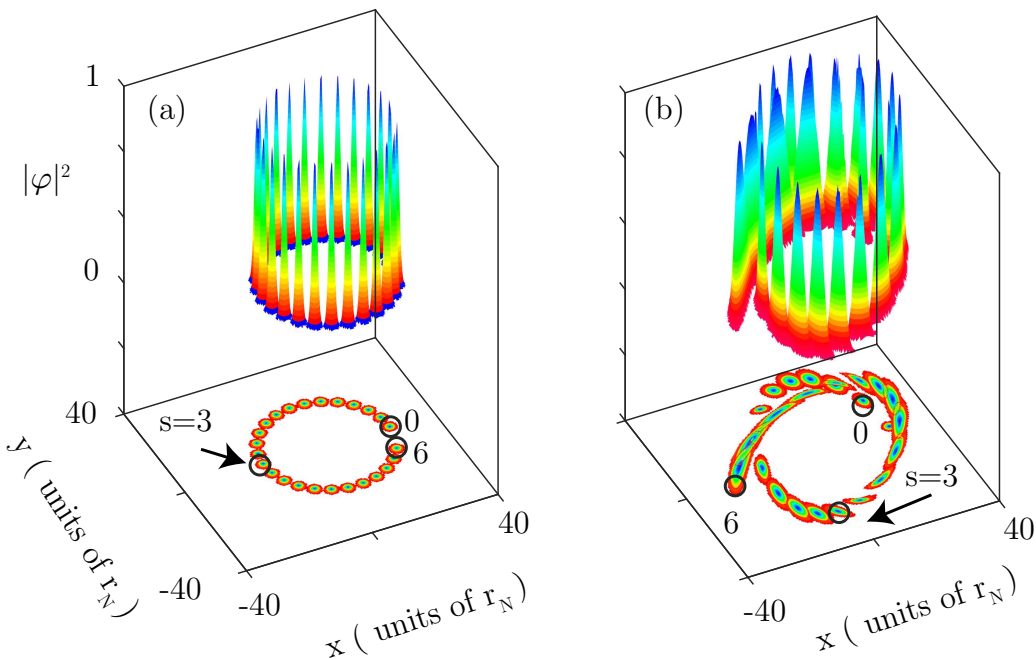}
\caption{
(Color online)  (a) Snapshots of $|\varphi|^2$ at different instants $s=t/t_N$ in the first orbit when $\varepsilon=0$;   (b) as in (a), for $\beta>0$ ($\varepsilon=0.01$). The panels show a two-dimensional visualization of $|\varphi|^2$ normalized to the maximal value. A precession occurs in the presence of the quantum gravity perturbation (parameters $u_o=20$, $w_o=0.1$, $k_l=20$, longer evolution is shown in Fig.~\ref{figure3D}).
\label{fourpanels}}
\end{figure*}
\begin{figure*}
\includegraphics[width=1.6\columnwidth]{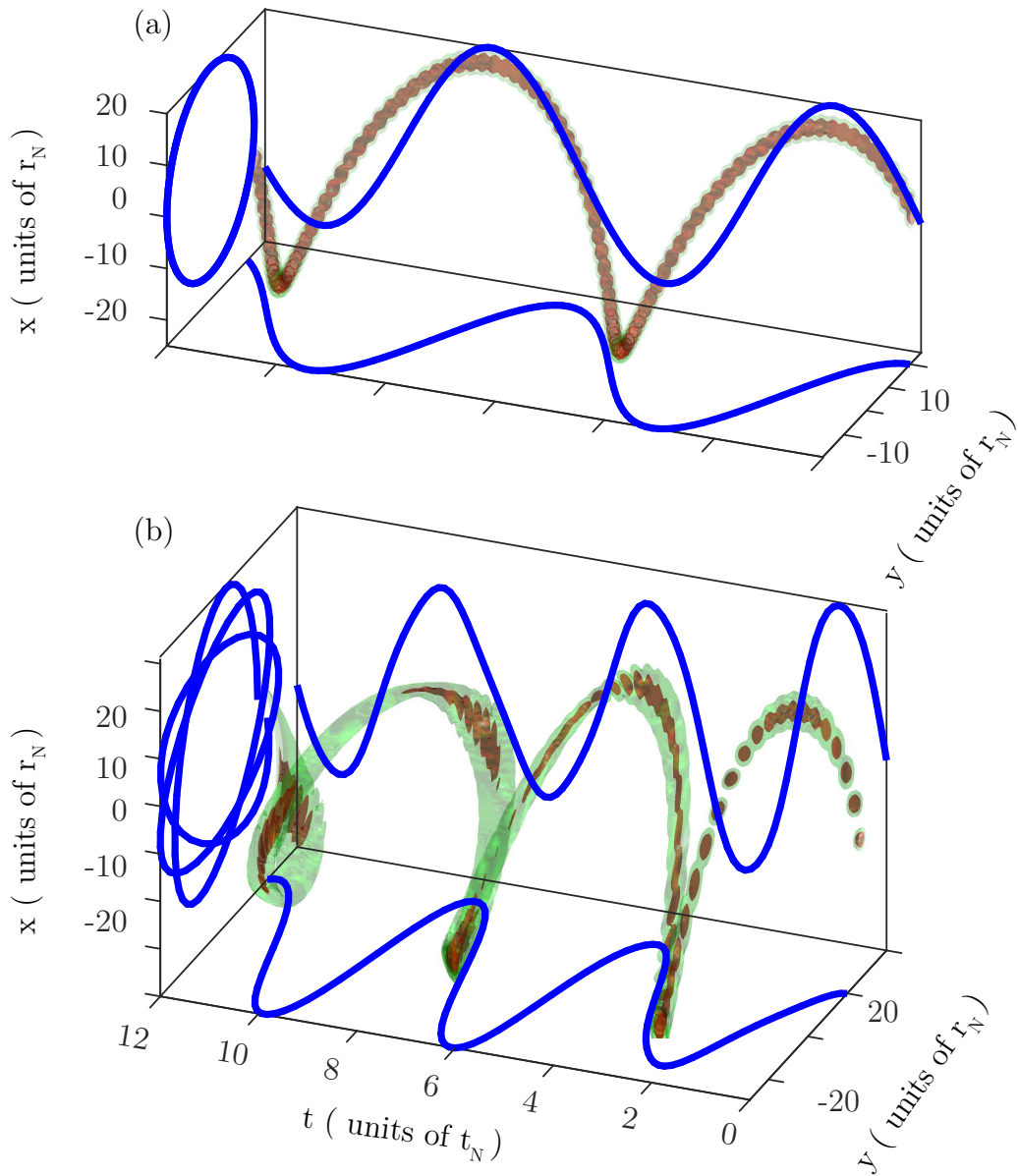}
\caption{ (Color online) (a)  Isosurface of the quantum fluid wave-packet orbiting in the
  parabolic potential in the absence of GUP effects ($\varepsilon=0.00$);
  (b)  as in (a) in the presence of GUP effects ($\varepsilon=0.01$).
  The panels show orthographic projections of the center-of-mass trajectories determined by the wave-function  (parameters $u_o=20$, $w_o=0.1$, $k_l=20$).
  \label{figure3D}}
\end{figure*}

\section{Numerical solution of the Schr\"odinger equation}
We validate our theoretical analysis on the generalized SE Eq.~(\ref{eq1}) in two dimensions.
We adopt dimensionless coordinates $s=t/t_N$, $u=x/w_N$, $v=y/w_N$,
with $w_N^2=\hbar t_N/m$, and $t_N=\Omega^{-1}$ and we have from Eq.~(\ref{eq1})
\begin{equation}
\imath \partial_s\varphi=\frac{1}{2}\nabla^2_{uv}\varphi+\frac{\varepsilon}{8}\nabla_{uv}^4\varphi+
\frac{1}{2}(u^2+v^2)\varphi\text{,}
\label{normalizedSE}
\end{equation}
with $\nabla^2_{u,v}=\partial_u^2+\partial_v^2$, and $\varepsilon=8 m \Omega \beta\hbar$. $\varepsilon=0$ corresponds to the absence of GUP effects. At $t=s=0$, a Gaussian wave-packet with waist $w_o$ is the initial condition:
\begin{equation}
  \varphi(u,v,s=0)=\varphi_o e^{-\left(\frac{u-u_o}{{\sqrt{2}w_o}}\right)^2
    -\left(\frac{v}{{\sqrt{2}w_o}}\right)^2} e^{\imath k_l v }\text{.}
  \label{initial}
\end{equation}
In (\ref{initial}), the angular momentum is $l= \hbar k_l u_0$.
\\If $\varepsilon=0$ and $k_l=0$, the wave-packet oscillates without orbiting
in the $y-$direction (not reported). Fig.~\ref{fourpanels}a shows the evolution, for $\varepsilon=0$ and $k_l=20$, by various snapshots of $|\varphi|^2$: the bottom panel reveals the elliptical orbit.
\\When $\varepsilon=0.01$, as in Fig.~\ref{fourpanels}b, we have evidence of the predicted precession. Figure~\ref{fourpanels} shows representative simulations, similar results occurs for all the considered cases.
\\Figure~\ref{figure3D}a shows a volumetric visualization of the solution of Eq.~(\ref{normalizedSE}) for a longer timescale with respect to Fig.~\ref{fourpanels}, for $\beta=\varepsilon=0$. Panels in Fig.~\ref{figure3D} give the trajectories of the wave-packet center of mass, which, for $\varepsilon=0$ do not reveal any precession. Figure~\ref{figure3D}b for $\varepsilon>0$ demonstrates the precession and the reduction of the orbital period.

\section{Precession in analogs and real experiments with quantum fluids}
\noindent To analyze possible experimental tests,
we recall that $\beta$ is  typically expressed in terms
of the dimensionless  
\begin{equation}\beta_0=\beta \frac{\hbar^2}{\ell_p^2}=\beta M_P^2 c^2\text{,}\end{equation} with $M_P$ the Planck mass, and $c$ the
vacuum light velocity. According to some authors,  $\beta_0< 10^{34}$ that we adopt as an optimistic upper bound for a real test of GUP phenomenology.~\cite{Das08}
For emulations, e.g., by paraxial light, we have $\beta_0=10^{55}$.~\cite{Conti2014,Braidotti201734}

We consider a wave-packet at initial distance $r_0$ from the center of the potential with tangential velocity $v_0$ and $|l|=m v_0 r_0=p_0 r_0$. For a harmonic trap, the quantum fluid size is  $\ell_B^2\cong \hbar/m\Omega$,~\cite{Dalfovo99} and we take $r_0\cong \ell_B$. A key point here is that the precession angle and the delay with respect to the
unperturbed case ($\beta=0$) increase at any orbit. Hence, the longer
the observation time, the more accessible the measurement of the
predicted perturbation. For a $1$ degree precession, the number of orbits is $\mathcal{N}_1=\pi/(180 \Delta)=\left(\frac{M_P c}{m v_0}\right)^2/(1440 \beta_0)$.
With $\beta_0=10^{34}$, $\ell_B=100\mu$m, and $v_0=10^{-10}c$,
for $^{87}$Rb BEC, we have $\mathcal{N}_1\simeq 10^{18}$ orbits occurring in a time interval $\mathcal{N}_1 T_0\simeq 10^{15}$~s, which is experimentally inaccessible.
On the contrary, if we consider a photonic condensate simulation, with $\beta_0=10^{55}$, $m=10^{-36}$kg, and $v_0=c$,~\cite{Klaers2010} we have $\mathcal{N}_1=0.03$: an experimentally testable 30 degrees precession in $T_0=2$ps.

Very interesting is the measurement of the time delay.
For example, one can measure the delay of the orbital oscillation with respect to a metrological reference clock. The period shrinks at any orbit by a relative amount $\delta T=-4\beta_0\left(\frac{mv_0}{M_P c}\right)^2$. For $^{87}$Rb BEC, we have $\delta T\simeq -10^{-20}$: after $10^6$ orbits ($15$ hours of observation), one has a delay of $1$fs.
The bounds for the modifications of the uncertainty principle become more precise when increasing the observation time. In the photonic simulation, one has $0.1$~ps delay for $1$ orbit.

We remark that in order to observe the precession, one has to generate states with an initial angular momentum in a trapping potential, as in Eq.~(\ref{initial}).
Considering the fact that our model applies to many physical systems, as in photonics, polaritonics, and Bose-Einstein condensates of atoms and photons, we remark that different approaches to realize experiments may be taken into account.

The literature and the experimental realizations of beams and condensates with angular momentum is so vast that cannot be reviewed here (see for example \cite{Padgett2011}). We will discuss in the following some representative cases. 

{\it For optical propagation}, a trapping transverse parabolic potential is realized by graded index systems, as lenses or optical fibers. A further possibility is to consider an array of optical lenses, which, as shown in \cite{YarivBook}, may also emulate a parabolic medium. 

In these devices, the initial state with angular momentum  in Eq.~(\ref{initial}) is excited by a beam spatially displaced with respect to the center of the trapping potential with an initial phase tilt.
The angular momentum is varied by the incidence angle with respect to the input plane. The beam follows a motion described by an optical orbit as represented in our theoretical analysis and in the simulations in figure \ref{figure3D}.
As a representative case, one can consider a beam with a transverse size of the order of few microns, with $1~\mu$m wavelength. The propagation length in fibers can be kilometers, this may enable a precise monitoring of the spiraling trajectories, and the deviation from an elliptical orbit due to the precession is here predicted.

{\it For atomic BEC}, states with angular momentum have been reported in a large number of articles (the interested readers may consider the references in \cite{Padgett2011}). The preparation of the initial state is conceptually similar to the optical case discussed above.
In a wide trapping potential, a fraction of condensed atoms is launched with an angular momentum parallel to the trap axis. This results into a spiraling motion of the atoms. An interesting possibility for putting atom into rotation is using transfer of orbital angular momentum from photons by stimulated Raman processes with Laguerre-Gauss optical beams, as, e.g., analysed in \cite{Helmerson2007}. This approach creates persistent currents in superfluid Bose gases, and the observation of precession dynamics in these systems may provide evidence of the effects discussed in this manuscript.

The case of {\it polaritonic condensates} is particularly relevant, as the generation of states with angular momentum has been actively investigated in recent years. For example, authors in \cite{Dall2014} demonstrated that angular momentum can be transferred to an exciton-polariton Bose-Einstein condensate by an external inchorent pump. In a parabolic potential, which is always present in this class of condensates, one can generate a rotating motion, and correspondingly observe the predicted precession.
Recent developments \cite{Bramati2016} show that it is possible to precisely control the amount of optical angular momentum,
and the generation of various spinning states. These results show that the observation of the precession here predicted is within current technologies for polariton superfluids.

As further possible experimental framework, we mention the case of photonic Bose-Einstein condensates. \cite{weitz2010} We are not aware of the experimental generation of photonic BEC with angular momentum. One can however figure out approaches similar to those discussed above for polaritonic BEC. Angular momentum may be transferred by an external pump to the condensate. A futher possibility is adopting symmetry breaking microcavities (e.g. by axicon mirrors) to generate spinning photonic BECs. The detailed analysis of these possibilities is beyond the scope of this manuscript.

\section{Conclusions}
In conclusion, the longtime observation of an orbiting quantum fluid is a feasible experimental road to set limits to hypothetical Planckian corrections to the known fundamental physical laws. 
We predict that the orbit of a wave-packet in a trapping potential exhibits a precession with an intriguing connection with the well-known anomalous
precession of the perihelion of Mercury, the first experimental test of general relativity.
In our case, the precession is due to the modification of the uncertainty principle predicted by the most studied theories of quantum gravity. The analogy with the original Einstein's solutions addresses the existence of additional effective quantum forces occurring at the Planck scale.

With reference to feasible real laboratory tests of quantum gravity theories, if one considers the optimistic estimate $\beta_0=10^{34}$,
the very small perturbation to the orbital period duration may become accessible after a large number of orbits because
of the cumulative amplification during time.
The ``true'' value of $\beta_0$ is unknown, and experiments with quantum fluids seem a concrete possibility to set bounds for new theories at the Planck scale. A metrological measurement of the period of a long-living BEC may unveil quantum gravity phenomenology. Very intriguing is adopting the Bose-Einstein condensates realized in the International Space Station at the NASA cold atom laboratory (CAL).~\cite{NASACAL}

With reference to laboratory analogs of the physics at the Planck scale, i.e., to the investigation of physical systems obeying laws mathematically identical to those supposed to be valid for quantum gravity, one can have $\beta_0\simeq 10^{55}$.
Correspondingly, quantum simulations and experiments with non-paraxial light, or polaritonic condensates, are well within current experimental possibilities. By quantum simulations, one may test the mathematical models, and also conceive improved frameworks for real probes of Planckian physics. This is specifically relevant for theories based on non-commutative coordinates,~\cite{Connes} for which experiments are lacking. 

Our results establish a bridge between quantum simulations and true experimental tests, and may be extended to other ``quantum gravity inspired'' effects, also including nonlinearity. These will be addressed in future works.
\section{Acknowledgments}
This publication was made possible through the support
of a grant from the John Templeton Foundation (grant
number 58277). We also acknowledge support the H2020 QuantERA Quomplex
(grant number 731743) and H2020 FET Project PhoQus.

\begin{thebibliography}{76}%
\makeatletter
\providecommand \@ifxundefined [1]{%
 \@ifx{#1\undefined}
}%
\providecommand \@ifnum [1]{%
 \ifnum #1\expandafter \@firstoftwo
 \else \expandafter \@secondoftwo
 \fi
}%
\providecommand \@ifx [1]{%
 \ifx #1\expandafter \@firstoftwo
 \else \expandafter \@secondoftwo
 \fi
}%
\providecommand \natexlab [1]{#1}%
\providecommand \enquote  [1]{``#1''}%
\providecommand \bibnamefont  [1]{#1}%
\providecommand \bibfnamefont [1]{#1}%
\providecommand \citenamefont [1]{#1}%
\providecommand \href@noop [0]{\@secondoftwo}%
\providecommand \href [0]{\begingroup \@sanitize@url \@href}%
\providecommand \@href[1]{\@@startlink{#1}\@@href}%
\providecommand \@@href[1]{\endgroup#1\@@endlink}%
\providecommand \@sanitize@url [0]{\catcode `\\12\catcode `\$12\catcode
  `\&12\catcode `\#12\catcode `\^12\catcode `\_12\catcode `\%12\relax}%
\providecommand \@@startlink[1]{}%
\providecommand \@@endlink[0]{}%
\providecommand \url  [0]{\begingroup\@sanitize@url \@url }%
\providecommand \@url [1]{\endgroup\@href {#1}{\urlprefix }}%
\providecommand \urlprefix  [0]{URL }%
\providecommand \Eprint [0]{\href }%
\providecommand \doibase [0]{http://dx.doi.org/}%
\providecommand \selectlanguage [0]{\@gobble}%
\providecommand \bibinfo  [0]{\@secondoftwo}%
\providecommand \bibfield  [0]{\@secondoftwo}%
\providecommand \translation [1]{[#1]}%
\providecommand \BibitemOpen [0]{}%
\providecommand \bibitemStop [0]{}%
\providecommand \bibitemNoStop [0]{.\EOS\space}%
\providecommand \EOS [0]{\spacefactor3000\relax}%
\providecommand \BibitemShut  [1]{\csname bibitem#1\endcsname}%
\let\auto@bib@innerbib\@empty
\bibitem [{\citenamefont {Hossenfelder}(2013)}]{Hossenfelder2013}%
  \BibitemOpen
  \bibfield  {author} {\bibinfo {author} {\bibfnamefont {S.}~\bibnamefont
  {Hossenfelder}},\ }\href {\doibase 10.12942/lrr-2013-2} {\bibfield  {journal}
  {\bibinfo  {journal} {Living Rev. Relativ.}\ }\textbf {\bibinfo {volume}
  {16}},\ \bibinfo {pages} {2} (\bibinfo {year} {2013})}\BibitemShut {NoStop}%
\bibitem [{\citenamefont {Amelino-Camelia}(2001)}]{AMELINOCAMELIA2001255}%
  \BibitemOpen
  \bibfield  {author} {\bibinfo {author} {\bibfnamefont {G.}~\bibnamefont
  {Amelino-Camelia}},\ }\href {\doibase
  https://doi.org/10.1016/S0370-2693(01)00506-8} {\bibfield  {journal}
  {\bibinfo  {journal} {Phys. Lett. B}\ }\textbf {\bibinfo {volume} {510}},\
  \bibinfo {pages} {255 } (\bibinfo {year} {2001})}\BibitemShut {NoStop}%
\bibitem [{\citenamefont {Gambini}\ and\ \citenamefont
  {Pullin}(1999)}]{Gambini99}%
  \BibitemOpen
  \bibfield  {author} {\bibinfo {author} {\bibfnamefont {R.}~\bibnamefont
  {Gambini}}\ and\ \bibinfo {author} {\bibfnamefont {J.}~\bibnamefont
  {Pullin}},\ }\href {\doibase 10.1103/PhysRevD.59.124021} {\bibfield
  {journal} {\bibinfo  {journal} {Phys.Rev. D}\ }\textbf {\bibinfo {volume}
  {59}},\ \bibinfo {pages} {124021} (\bibinfo {year} {1999})}\BibitemShut
  {NoStop}%
\bibitem [{\citenamefont {Mukhanov}\ and\ \citenamefont
  {Winitzki}(2007)}]{MukhanovBook2007}%
  \BibitemOpen
  \bibfield  {author} {\bibinfo {author} {\bibfnamefont {V.~F.}\ \bibnamefont
  {Mukhanov}}\ and\ \bibinfo {author} {\bibfnamefont {S.}~\bibnamefont
  {Winitzki}},\ }\href@noop {} {\emph {\bibinfo {title} {Quantum Effects in
  Gravity}}}\ (\bibinfo  {publisher} {Cambridge University Press},\ \bibinfo
  {year} {2007})\BibitemShut {NoStop}%
\bibitem [{\citenamefont {Rovelli}\ and\ \citenamefont
  {Vidotto}(2015)}]{BookRovelli2015}%
  \BibitemOpen
  \bibfield  {author} {\bibinfo {author} {\bibfnamefont {C.}~\bibnamefont
  {Rovelli}}\ and\ \bibinfo {author} {\bibfnamefont {F.}~\bibnamefont
  {Vidotto}},\ }\href@noop {} {\emph {\bibinfo {title} {Covariant Loop Quantum
  Gravity}}}\ (\bibinfo  {publisher} {Cambridge University Press},\ \bibinfo
  {year} {2015})\BibitemShut {NoStop}%
\bibitem [{\citenamefont {Thiemann}(2007)}]{ThiemannBook2007}%
  \BibitemOpen
  \bibfield  {author} {\bibinfo {author} {\bibfnamefont {T.}~\bibnamefont
  {Thiemann}},\ }\href@noop {} {\emph {\bibinfo {title} {Modern Canonical
  Quantum General Relativity}}}\ (\bibinfo  {publisher} {Cambridge University
  Press},\ \bibinfo {year} {2007})\BibitemShut {NoStop}%
\bibitem [{\citenamefont {Kiefer}(2012)}]{BookKiefer2012}%
  \BibitemOpen
  \bibfield  {author} {\bibinfo {author} {\bibfnamefont {C.}~\bibnamefont
  {Kiefer}},\ }\href@noop {} {\emph {\bibinfo {title} {Quantum Gravity}}}\
  (\bibinfo  {publisher} {Oxford University Press},\ \bibinfo {year}
  {2012})\BibitemShut {NoStop}%
\bibitem [{\citenamefont {Fedi}()}]{Fedi2018}%
  \BibitemOpen
  \bibfield  {author} {\bibinfo {author} {\bibfnamefont {M.}~\bibnamefont
  {Fedi}},\ }\href@noop {} {\bibinfo  {journal} {hal-01648358v5}\ }\BibitemShut
  {NoStop}%
\bibitem [{\citenamefont {Maselli}\ \emph {et~al.}(2018)\citenamefont
  {Maselli}, \citenamefont {Pani}, \citenamefont {Cardoso}, \citenamefont
  {Abdelsalhin}, \citenamefont {Gualtieri},\ and\ \citenamefont
  {Ferrari}}]{PhysRevLett.120.081101}%
  \BibitemOpen
\bibfield  {journal} {  }\bibfield  {author} {\bibinfo {author} {\bibfnamefont
  {A.}~\bibnamefont {Maselli}}, \bibinfo {author} {\bibfnamefont
  {P.}~\bibnamefont {Pani}}, \bibinfo {author} {\bibfnamefont {V.}~\bibnamefont
  {Cardoso}}, \bibinfo {author} {\bibfnamefont {T.}~\bibnamefont
  {Abdelsalhin}}, \bibinfo {author} {\bibfnamefont {L.}~\bibnamefont
  {Gualtieri}}, \ and\ \bibinfo {author} {\bibfnamefont {V.}~\bibnamefont
  {Ferrari}},\ }\href {\doibase 10.1103/PhysRevLett.120.081101} {\bibfield
  {journal} {\bibinfo  {journal} {Phys. Rev. Lett.}\ }\textbf {\bibinfo
  {volume} {120}},\ \bibinfo {pages} {081101} (\bibinfo {year}
  {2018})}\BibitemShut {NoStop}%
\bibitem [{\citenamefont {Garay}\ \emph {et~al.}(2000)\citenamefont {Garay},
  \citenamefont {Anglin}, \citenamefont {Cirac},\ and\ \citenamefont
  {Zoller}}]{PhysRevLett.85.4643}%
  \BibitemOpen
  \bibfield  {author} {\bibinfo {author} {\bibfnamefont {L.~J.}\ \bibnamefont
  {Garay}}, \bibinfo {author} {\bibfnamefont {J.~R.}\ \bibnamefont {Anglin}},
  \bibinfo {author} {\bibfnamefont {J.~I.}\ \bibnamefont {Cirac}}, \ and\
  \bibinfo {author} {\bibfnamefont {P.}~\bibnamefont {Zoller}},\ }\href
  {\doibase 10.1103/PhysRevLett.85.4643} {\bibfield  {journal} {\bibinfo
  {journal} {Phys. Rev. Lett.}\ }\textbf {\bibinfo {volume} {85}},\ \bibinfo
  {pages} {4643} (\bibinfo {year} {2000})}\BibitemShut {NoStop}%
\bibitem [{\citenamefont {Marino}\ \emph {et~al.}(2009)\citenamefont {Marino},
  \citenamefont {Ciszak},\ and\ \citenamefont {Ortolan}}]{PhysRevA.80.065802}%
  \BibitemOpen
  \bibfield  {author} {\bibinfo {author} {\bibfnamefont {F.}~\bibnamefont
  {Marino}}, \bibinfo {author} {\bibfnamefont {M.}~\bibnamefont {Ciszak}}, \
  and\ \bibinfo {author} {\bibfnamefont {A.}~\bibnamefont {Ortolan}},\ }\href
  {\doibase 10.1103/PhysRevA.80.065802} {\bibfield  {journal} {\bibinfo
  {journal} {Phys. Rev. A}\ }\textbf {\bibinfo {volume} {80}},\ \bibinfo
  {pages} {065802} (\bibinfo {year} {2009})}\BibitemShut {NoStop}%
\bibitem [{\citenamefont {{Vocke}}\ \emph {et~al.}(2017)\citenamefont
  {{Vocke}}, \citenamefont {{Maitland}}, \citenamefont {{Prain}}, \citenamefont
  {{Biancalana}}, \citenamefont {{Marino}}, \citenamefont {{Wright}},\ and\
  \citenamefont {{Faccio}}}]{Faccio2017}%
  \BibitemOpen
  \bibfield  {author} {\bibinfo {author} {\bibfnamefont {D.}~\bibnamefont
  {{Vocke}}}, \bibinfo {author} {\bibfnamefont {C.}~\bibnamefont {{Maitland}}},
  \bibinfo {author} {\bibfnamefont {A.}~\bibnamefont {{Prain}}}, \bibinfo
  {author} {\bibfnamefont {F.}~\bibnamefont {{Biancalana}}}, \bibinfo {author}
  {\bibfnamefont {F.}~\bibnamefont {{Marino}}}, \bibinfo {author}
  {\bibfnamefont {E.~M.}\ \bibnamefont {{Wright}}}, \ and\ \bibinfo {author}
  {\bibfnamefont {D.}~\bibnamefont {{Faccio}}},\ }\href@noop {} {\bibfield
  {journal} {\bibinfo  {journal} {ArXiv e-prints}\ } (\bibinfo {year}
  {2017})},\ \Eprint {http://arxiv.org/abs/1709.04293} {arXiv:1709.04293}
  \BibitemShut {NoStop}%
\bibitem [{\citenamefont {Steinhauer}(2016)}]{Steinhauer2016}%
  \BibitemOpen
  \bibfield  {author} {\bibinfo {author} {\bibfnamefont {J.}~\bibnamefont
  {Steinhauer}},\ }\href {http://dx.doi.org/10.1038/nphys3863} {\bibfield
  {journal} {\bibinfo  {journal} {Nat. Phys.}\ }\textbf {\bibinfo {volume}
  {12}},\ \bibinfo {pages} {959} (\bibinfo {year} {2016})}\BibitemShut
  {NoStop}%
\bibitem [{\citenamefont {Faccio}(2012)}]{faccio2012}%
  \BibitemOpen
  \bibfield  {author} {\bibinfo {author} {\bibfnamefont {D.}~\bibnamefont
  {Faccio}},\ }\href {\doibase 10.1080/00107514.2011.642559} {\bibfield
  {journal} {\bibinfo  {journal} {Contemp. Phys.}\ }\textbf {\bibinfo {volume}
  {53}},\ \bibinfo {pages} {97–112} (\bibinfo {year} {2012})}\BibitemShut
  {NoStop}%
\bibitem [{\citenamefont {Unruh}(1976)}]{unruh1976}%
  \BibitemOpen
  \bibfield  {author} {\bibinfo {author} {\bibfnamefont {W.~G.}\ \bibnamefont
  {Unruh}},\ }\href {\doibase 10.1103/PhysRevD.14.870} {\bibfield  {journal}
  {\bibinfo  {journal} {Phys. Rev. D}\ }\textbf {\bibinfo {volume} {14}},\
  \bibinfo {pages} {870–892} (\bibinfo {year} {1976})}\BibitemShut {NoStop}%
\bibitem [{\citenamefont {Ornigotti}\ \emph {et~al.}(2018)\citenamefont
  {Ornigotti}, \citenamefont {Bar-Ad}, \citenamefont {Szameit},\ and\
  \citenamefont {Fleurov}}]{PhysRevA.97.013823}%
  \BibitemOpen
  \bibfield  {author} {\bibinfo {author} {\bibfnamefont {M.}~\bibnamefont
  {Ornigotti}}, \bibinfo {author} {\bibfnamefont {S.}~\bibnamefont {Bar-Ad}},
  \bibinfo {author} {\bibfnamefont {A.}~\bibnamefont {Szameit}}, \ and\
  \bibinfo {author} {\bibfnamefont {V.}~\bibnamefont {Fleurov}},\ }\href
  {\doibase 10.1103/PhysRevA.97.013823} {\bibfield  {journal} {\bibinfo
  {journal} {Phys. Rev. A}\ }\textbf {\bibinfo {volume} {97}},\ \bibinfo
  {pages} {013823} (\bibinfo {year} {2018})}\BibitemShut {NoStop}%
\bibitem [{\citenamefont {Ch\"a}\ and\ \citenamefont
  {Fischer}(2017)}]{Fischer2017}%
  \BibitemOpen
  \bibfield  {author} {\bibinfo {author} {\bibfnamefont {S.-Y.}\ \bibnamefont
  {Ch\"a}}\ and\ \bibinfo {author} {\bibfnamefont {U.~R.}\ \bibnamefont
  {Fischer}},\ }\href {\doibase 10.1103/PhysRevLett.118.130404} {\bibfield
  {journal} {\bibinfo  {journal} {Phys. Rev. Lett.}\ }\textbf {\bibinfo
  {volume} {118}},\ \bibinfo {pages} {130404} (\bibinfo {year}
  {2017})}\BibitemShut {NoStop}%
\bibitem [{\citenamefont {Eckel}\ \emph {et~al.}(2018)\citenamefont {Eckel},
  \citenamefont {Kumar}, \citenamefont {Jacobson}, \citenamefont {Spielman},\
  and\ \citenamefont {Campbell}}]{Eckel2018}%
  \BibitemOpen
  \bibfield  {author} {\bibinfo {author} {\bibfnamefont {S.}~\bibnamefont
  {Eckel}}, \bibinfo {author} {\bibfnamefont {A.}~\bibnamefont {Kumar}},
  \bibinfo {author} {\bibfnamefont {T.}~\bibnamefont {Jacobson}}, \bibinfo
  {author} {\bibfnamefont {I.~B.}\ \bibnamefont {Spielman}}, \ and\ \bibinfo
  {author} {\bibfnamefont {G.~K.}\ \bibnamefont {Campbell}},\ }\href {\doibase
  10.1103/PhysRevX.8.021021} {\bibfield  {journal} {\bibinfo  {journal} {Phys.
  Rev. X}\ }\textbf {\bibinfo {volume} {8}},\ \bibinfo {pages} {021021}
  (\bibinfo {year} {2018})}\BibitemShut {NoStop}%
\bibitem [{\citenamefont {Paredes}\ and\ \citenamefont
  {Michinel}(2016)}]{Paredes2016}%
  \BibitemOpen
  \bibfield  {author} {\bibinfo {author} {\bibfnamefont {A.}~\bibnamefont
  {Paredes}}\ and\ \bibinfo {author} {\bibfnamefont {H.}~\bibnamefont
  {Michinel}},\ }\href {\doibase https://doi.org/10.1016/j.dark.2016.02.003}
  {\bibfield  {journal} {\bibinfo  {journal} {Phys. Dark Universe}\ }\textbf
  {\bibinfo {volume} {12}},\ \bibinfo {pages} {50 } (\bibinfo {year}
  {2016})}\BibitemShut {NoStop}%
\bibitem [{\citenamefont {Zeng}\ \emph {et~al.}(2006)\citenamefont {Zeng},
  \citenamefont {Wu}, \citenamefont {Xu},\ and\ \citenamefont {Wu}}]{Zheng06}%
  \BibitemOpen
  \bibfield  {author} {\bibinfo {author} {\bibfnamefont {H.}~\bibnamefont
  {Zeng}}, \bibinfo {author} {\bibfnamefont {J.}~\bibnamefont {Wu}}, \bibinfo
  {author} {\bibfnamefont {H.}~\bibnamefont {Xu}}, \ and\ \bibinfo {author}
  {\bibfnamefont {K.}~\bibnamefont {Wu}},\ }\href
  {http://link.aps.org/abstract/PRL/v96/e083902} {\bibfield  {journal}
  {\bibinfo  {journal} {Phys. Rev. Lett.}\ }\textbf {\bibinfo {volume} {96}},\
  \bibinfo {eid} {083902} (\bibinfo {year} {2006})}\BibitemShut {NoStop}%
\bibitem [{\citenamefont {Barcelo}\ \emph {et~al.}(2003)\citenamefont
  {Barcelo}, \citenamefont {Liberati},\ and\ \citenamefont
  {Visser}}]{Barcelo2003}%
  \BibitemOpen
  \bibfield  {author} {\bibinfo {author} {\bibfnamefont {C.}~\bibnamefont
  {Barcelo}}, \bibinfo {author} {\bibfnamefont {S.}~\bibnamefont {Liberati}}, \
  and\ \bibinfo {author} {\bibfnamefont {M.}~\bibnamefont {Visser}},\ }\href
  {\doibase 10.1103/PhysRevA.68.053613} {\bibfield  {journal} {\bibinfo
  {journal} {Phys. Rev. A}\ }\textbf {\bibinfo {volume} {68}},\ \bibinfo
  {pages} {053613} (\bibinfo {year} {2003})}\BibitemShut {NoStop}%
\bibitem [{\citenamefont {Longhi}(2011)}]{Longhi2011}%
  \BibitemOpen
  \bibfield  {author} {\bibinfo {author} {\bibfnamefont {S.}~\bibnamefont
  {Longhi}},\ }\href {\doibase 10.1007/s00340-011-4628-7} {\bibfield  {journal}
  {\bibinfo  {journal} {Appl. Phys, B}\ }\textbf {\bibinfo {volume} {104}},\
  \bibinfo {pages} {453–468} (\bibinfo {year} {2011})}\BibitemShut {NoStop}%
\bibitem [{\citenamefont {Braidotti}\ and\ \citenamefont
  {Conti}(2017)}]{Braidotti2017}%
  \BibitemOpen
  \bibfield  {author} {\bibinfo {author} {\bibfnamefont {M.~C.}\ \bibnamefont
  {Braidotti}}\ and\ \bibinfo {author} {\bibfnamefont {C.}~\bibnamefont
  {Conti}},\ }\href@noop {} {\bibfield  {journal} {\bibinfo  {journal} {ArXiv
  e-prints}\ } (\bibinfo {year} {2017})},\ \Eprint
  {http://arxiv.org/abs/1708.02623} {arXiv:1708.02623} \BibitemShut {NoStop}%
\bibitem [{\citenamefont {Gentilini}\ \emph {et~al.}(2015)\citenamefont
  {Gentilini}, \citenamefont {Braidotti}, \citenamefont {Marcucci},
  \citenamefont {DelRe},\ and\ \citenamefont {Conti}}]{Gentilini2015Glauber}%
  \BibitemOpen
  \bibfield  {author} {\bibinfo {author} {\bibfnamefont {S.}~\bibnamefont
  {Gentilini}}, \bibinfo {author} {\bibfnamefont {M.~C.}\ \bibnamefont
  {Braidotti}}, \bibinfo {author} {\bibfnamefont {G.}~\bibnamefont {Marcucci}},
  \bibinfo {author} {\bibfnamefont {E.}~\bibnamefont {DelRe}}, \ and\ \bibinfo
  {author} {\bibfnamefont {C.}~\bibnamefont {Conti}},\ }\href {\doibase
  10.1038/srep15816} {\bibfield  {journal} {\bibinfo  {journal} {Sci. Rep.}\
  }\textbf {\bibinfo {volume} {5}},\ \bibinfo {pages} {15816} (\bibinfo {year}
  {2015})}\BibitemShut {NoStop}%
\bibitem [{\citenamefont {Marcucci}\ and\ \citenamefont
  {Conti}(2016)}]{PhysRevA.94.052136}%
  \BibitemOpen
  \bibfield  {author} {\bibinfo {author} {\bibfnamefont {G.}~\bibnamefont
  {Marcucci}}\ and\ \bibinfo {author} {\bibfnamefont {C.}~\bibnamefont
  {Conti}},\ }\href {\doibase 10.1103/PhysRevA.94.052136} {\bibfield  {journal}
  {\bibinfo  {journal} {Phys. Rev. A}\ }\textbf {\bibinfo {volume} {94}},\
  \bibinfo {pages} {052136} (\bibinfo {year} {2016})}\BibitemShut {NoStop}%
\bibitem [{\citenamefont {Bohm}(1999)}]{PhysRevA.60.861}%
  \BibitemOpen
  \bibfield  {author} {\bibinfo {author} {\bibfnamefont {A.}~\bibnamefont
  {Bohm}},\ }\href {\doibase 10.1103/PhysRevA.60.861} {\bibfield  {journal}
  {\bibinfo  {journal} {Phys. Rev. A}\ }\textbf {\bibinfo {volume} {60}},\
  \bibinfo {pages} {861} (\bibinfo {year} {1999})}\BibitemShut {NoStop}%
\bibitem [{\citenamefont {Kempf}\ \emph {et~al.}(1995)\citenamefont {Kempf},
  \citenamefont {Mangano},\ and\ \citenamefont {Mann}}]{Kempf95}%
  \BibitemOpen
  \bibfield  {author} {\bibinfo {author} {\bibfnamefont {A.}~\bibnamefont
  {Kempf}}, \bibinfo {author} {\bibfnamefont {G.}~\bibnamefont {Mangano}}, \
  and\ \bibinfo {author} {\bibfnamefont {R.~B.}\ \bibnamefont {Mann}},\ }\href
  {\doibase 10.1103/PhysRevD.52.1108} {\bibfield  {journal} {\bibinfo
  {journal} {Phys.Rev.D}\ }\textbf {\bibinfo {volume} {52}},\ \bibinfo {pages}
  {1108–1118} (\bibinfo {year} {1995})}\BibitemShut {NoStop}%
\bibitem [{\citenamefont {Garay}(1995)}]{Garay1995}%
  \BibitemOpen
  \bibfield  {author} {\bibinfo {author} {\bibfnamefont {L.~J.}\ \bibnamefont
  {Garay}},\ }\href {\doibase 10.1142/S0217751X95000085} {\bibfield  {journal}
  {\bibinfo  {journal} {Int. J. Mod. Phys. A}\ }\textbf {\bibinfo {volume}
  {10}},\ \bibinfo {pages} {145} (\bibinfo {year} {1995})}\BibitemShut
  {NoStop}%
\bibitem [{\citenamefont {Kempf}\ and\ \citenamefont
  {Mangano}(1997)}]{KempfMangano97}%
  \BibitemOpen
  \bibfield  {author} {\bibinfo {author} {\bibfnamefont {A.}~\bibnamefont
  {Kempf}}\ and\ \bibinfo {author} {\bibfnamefont {G.}~\bibnamefont
  {Mangano}},\ }\href {\doibase 10.1103/PhysRevD.55.7909} {\bibfield  {journal}
  {\bibinfo  {journal} {Phys. Rev. D}\ }\textbf {\bibinfo {volume} {55}},\
  \bibinfo {pages} {7909} (\bibinfo {year} {1997})}\BibitemShut {NoStop}%
\bibitem [{\citenamefont {Witten}(1996)}]{Witten1996}%
  \BibitemOpen
  \bibfield  {author} {\bibinfo {author} {\bibfnamefont {E.}~\bibnamefont
  {Witten}},\ }\href@noop {} {\bibfield  {journal} {\bibinfo  {journal}
  {Physics Today}\ }\textbf {\bibinfo {volume} {49}},\ \bibinfo {pages} {24}
  (\bibinfo {year} {1996})}\BibitemShut {NoStop}%
\bibitem [{\citenamefont {Connes}(1994)}]{Connes}%
  \BibitemOpen
  \bibfield  {author} {\bibinfo {author} {\bibfnamefont {A.}~\bibnamefont
  {Connes}},\ }\href@noop {} {\emph {\bibinfo {title} {Noncommutative
  Geometry:}}}\ (\bibinfo  {publisher} {Academic Press},\ \bibinfo {year}
  {1994})\BibitemShut {NoStop}%
\bibitem [{\citenamefont {Maggiore}(1993)}]{Maggiore1993}%
  \BibitemOpen
  \bibfield  {author} {\bibinfo {author} {\bibfnamefont {M.}~\bibnamefont
  {Maggiore}},\ }\href {\doibase https://doi.org/10.1016/0370-2693(93)90785-G}
  {\bibfield  {journal} {\bibinfo  {journal} {Phys. Lett. B}\ }\textbf
  {\bibinfo {volume} {319}},\ \bibinfo {pages} {83 } (\bibinfo {year}
  {1993})}\BibitemShut {NoStop}%
\bibitem [{\citenamefont {Amelino-Camelia}(2013)}]{ACamelia13}%
  \BibitemOpen
  \bibfield  {author} {\bibinfo {author} {\bibfnamefont {G.}~\bibnamefont
  {Amelino-Camelia}},\ }\href {\doibase 10.1103/PhysRevLett.111.101301}
  {\bibfield  {journal} {\bibinfo  {journal} {Phys. Rev. Lett.}\ }\textbf
  {\bibinfo {volume} {111}},\ \bibinfo {pages} {101301} (\bibinfo {year}
  {2013})}\BibitemShut {NoStop}%
\bibitem [{\citenamefont {Douglas}\ and\ \citenamefont
  {Nekrasov}(2001)}]{Douglas2001}%
  \BibitemOpen
  \bibfield  {author} {\bibinfo {author} {\bibfnamefont {M.~R.}\ \bibnamefont
  {Douglas}}\ and\ \bibinfo {author} {\bibfnamefont {N.~A.}\ \bibnamefont
  {Nekrasov}},\ }\href {\doibase 10.1103/RevModPhys.73.977} {\bibfield
  {journal} {\bibinfo  {journal} {Rev. Mod. Phys.}\ }\textbf {\bibinfo {volume}
  {73}},\ \bibinfo {pages} {977} (\bibinfo {year} {2001})}\BibitemShut
  {NoStop}%
\bibitem [{\citenamefont {Ghosh}(2014)}]{Ghosh2014}%
  \BibitemOpen
  \bibfield  {author} {\bibinfo {author} {\bibfnamefont {S.}~\bibnamefont
  {Ghosh}},\ }\href {http://stacks.iop.org/0264-9381/31/i=2/a=025025}
  {\bibfield  {journal} {\bibinfo  {journal} {Classical Quantum Gravity}\
  }\textbf {\bibinfo {volume} {31}},\ \bibinfo {pages} {025025} (\bibinfo
  {year} {2014})}\BibitemShut {NoStop}%
\bibitem [{\citenamefont {Scardigli}(1999)}]{Scardigli99}%
  \BibitemOpen
  \bibfield  {author} {\bibinfo {author} {\bibfnamefont {F.}~\bibnamefont
  {Scardigli}},\ }\href {\doibase
  https://doi.org/10.1016/S0370-2693(99)00167-7} {\bibfield  {journal}
  {\bibinfo  {journal} {Physics Letters B}\ }\textbf {\bibinfo {volume}
  {452}},\ \bibinfo {pages} {39 } (\bibinfo {year} {1999})}\BibitemShut
  {NoStop}%
\bibitem [{\citenamefont {Chang}\ \emph {et~al.}(2002)\citenamefont {Chang},
  \citenamefont {Minic}, \citenamefont {Okamura},\ and\ \citenamefont
  {Takeuchi}}]{Chang02}%
  \BibitemOpen
  \bibfield  {author} {\bibinfo {author} {\bibfnamefont {L.~N.}\ \bibnamefont
  {Chang}}, \bibinfo {author} {\bibfnamefont {D.}~\bibnamefont {Minic}},
  \bibinfo {author} {\bibfnamefont {N.}~\bibnamefont {Okamura}}, \ and\
  \bibinfo {author} {\bibfnamefont {T.}~\bibnamefont {Takeuchi}},\ }\href
  {\doibase 10.1103/PhysRevD.65.125028} {\bibfield  {journal} {\bibinfo
  {journal} {Phys. Rev. D}\ }\textbf {\bibinfo {volume} {65}},\ \bibinfo
  {pages} {125028} (\bibinfo {year} {2002})}\BibitemShut {NoStop}%
\bibitem [{\citenamefont {Das}\ and\ \citenamefont {Vagenas}(2008)}]{Das08}%
  \BibitemOpen
  \bibfield  {author} {\bibinfo {author} {\bibfnamefont {S.}~\bibnamefont
  {Das}}\ and\ \bibinfo {author} {\bibfnamefont {E.~C.}\ \bibnamefont
  {Vagenas}},\ }\href {\doibase 10.1103/PhysRevLett.101.221301} {\bibfield
  {journal} {\bibinfo  {journal} {Phys.Rev.Lett.}\ }\textbf {\bibinfo {volume}
  {101}},\ \bibinfo {pages} {221301} (\bibinfo {year} {2008})}\BibitemShut
  {NoStop}%
\bibitem [{\citenamefont {Pedram}(2012)}]{Pedram12}%
  \BibitemOpen
  \bibfield  {author} {\bibinfo {author} {\bibfnamefont {P.}~\bibnamefont
  {Pedram}},\ }\href {\doibase 10.1016/j.physletb.2012.07.005} {\bibfield
  {journal} {\bibinfo  {journal} {Phys.Lett. B}\ }\textbf {\bibinfo {volume}
  {714}},\ \bibinfo {pages} {317–323} (\bibinfo {year} {2012})}\BibitemShut
  {NoStop}%
\bibitem [{\citenamefont {Pikovski}\ \emph {et~al.}(2012)\citenamefont
  {Pikovski}, \citenamefont {Vanner}, \citenamefont {Aspelmeyer}, \citenamefont
  {Kim},\ and\ \citenamefont {Brukner}}]{Pikovski12}%
  \BibitemOpen
  \bibfield  {author} {\bibinfo {author} {\bibfnamefont {I.}~\bibnamefont
  {Pikovski}}, \bibinfo {author} {\bibfnamefont {M.~R.}\ \bibnamefont
  {Vanner}}, \bibinfo {author} {\bibfnamefont {M.}~\bibnamefont {Aspelmeyer}},
  \bibinfo {author} {\bibfnamefont {M.~S.}\ \bibnamefont {Kim}}, \ and\
  \bibinfo {author} {\bibfnamefont {C.}~\bibnamefont {Brukner}},\ }\href
  {http://www.nature.com/nphys/journal/v8/n5/abs/nphys2262.html#supplementary-information}
  {\bibfield  {journal} {\bibinfo  {journal} {Nat. Phys.}\ }\textbf {\bibinfo
  {volume} {8}},\ \bibinfo {pages} {393–397} (\bibinfo {year}
  {2012})}\BibitemShut {NoStop}%
\bibitem [{\citenamefont {Bekenstein}(2012)}]{Bekenstein12}%
  \BibitemOpen
  \bibfield  {author} {\bibinfo {author} {\bibfnamefont {J.~D.}\ \bibnamefont
  {Bekenstein}},\ }\href {\doibase 10.1103/PhysRevD.86.124040} {\bibfield
  {journal} {\bibinfo  {journal} {Phys. Rev. D}\ }\textbf {\bibinfo {volume}
  {86}},\ \bibinfo {pages} {124040} (\bibinfo {year} {2012})}\BibitemShut
  {NoStop}%
\bibitem [{\citenamefont {Bawaj}\ \emph {et~al.}(2015)\citenamefont {Bawaj},
  \citenamefont {Biancofiore}, \citenamefont {Bonaldi}, \citenamefont
  {Bonfigli}, \citenamefont {Borrielli}, \citenamefont {Di~Giuseppe},
  \citenamefont {Marconi}, \citenamefont {Marino}, \citenamefont {Natali},
  \citenamefont {Pontin}, \citenamefont {Prodi}, \citenamefont {Serra},
  \citenamefont {Vitali},\ and\ \citenamefont {Marin}}]{Marin2015}%
  \BibitemOpen
  \bibfield  {author} {\bibinfo {author} {\bibfnamefont {M.}~\bibnamefont
  {Bawaj}}, \bibinfo {author} {\bibfnamefont {C.}~\bibnamefont {Biancofiore}},
  \bibinfo {author} {\bibfnamefont {M.}~\bibnamefont {Bonaldi}}, \bibinfo
  {author} {\bibfnamefont {F.}~\bibnamefont {Bonfigli}}, \bibinfo {author}
  {\bibfnamefont {A.}~\bibnamefont {Borrielli}}, \bibinfo {author}
  {\bibfnamefont {G.}~\bibnamefont {Di~Giuseppe}}, \bibinfo {author}
  {\bibfnamefont {L.}~\bibnamefont {Marconi}}, \bibinfo {author} {\bibfnamefont
  {F.}~\bibnamefont {Marino}}, \bibinfo {author} {\bibfnamefont
  {R.}~\bibnamefont {Natali}}, \bibinfo {author} {\bibfnamefont
  {A.}~\bibnamefont {Pontin}}, \bibinfo {author} {\bibfnamefont {G.~A.}\
  \bibnamefont {Prodi}}, \bibinfo {author} {\bibfnamefont {E.}~\bibnamefont
  {Serra}}, \bibinfo {author} {\bibfnamefont {D.}~\bibnamefont {Vitali}}, \
  and\ \bibinfo {author} {\bibfnamefont {F.}~\bibnamefont {Marin}},\ }\href
  {http://dx.doi.org/10.1038/ncomms8503} {\bibfield  {journal} {\bibinfo
  {journal} {Nat. Commun.}\ }\textbf {\bibinfo {volume} {6}},\ \bibinfo {pages}
  {7503} (\bibinfo {year} {2015})}\BibitemShut {NoStop}%
\bibitem [{\citenamefont {Silagadze}(2009)}]{Silagadze2009}%
  \BibitemOpen
  \bibfield  {author} {\bibinfo {author} {\bibfnamefont {Z.}~\bibnamefont
  {Silagadze}},\ }\href {\doibase
  https://doi.org/10.1016/j.physleta.2009.05.053} {\bibfield  {journal}
  {\bibinfo  {journal} {Phys. Lett. A}\ }\textbf {\bibinfo {volume} {373}},\
  \bibinfo {pages} {2643 } (\bibinfo {year} {2009})}\BibitemShut {NoStop}%
\bibitem [{\citenamefont {Pedram}\ \emph {et~al.}(2011)\citenamefont {Pedram},
  \citenamefont {Nozari},\ and\ \citenamefont {Taheri}}]{Pedram2011}%
  \BibitemOpen
  \bibfield  {author} {\bibinfo {author} {\bibfnamefont {P.}~\bibnamefont
  {Pedram}}, \bibinfo {author} {\bibfnamefont {K.}~\bibnamefont {Nozari}}, \
  and\ \bibinfo {author} {\bibfnamefont {S.~H.}\ \bibnamefont {Taheri}},\
  }\href {\doibase 10.1007/JHEP03(2011)093} {\bibfield  {journal} {\bibinfo
  {journal} {J. High Energy Phys.}\ }\textbf {\bibinfo {volume} {2011}},\
  \bibinfo {pages} {93} (\bibinfo {year} {2011})}\BibitemShut {NoStop}%
\bibitem [{\citenamefont {Conti}(2014)}]{Conti2014}%
  \BibitemOpen
  \bibfield  {author} {\bibinfo {author} {\bibfnamefont {C.}~\bibnamefont
  {Conti}},\ }\href {\doibase 10.1103/PhysRevA.89.061801} {\bibfield  {journal}
  {\bibinfo  {journal} {Phys. Rev. A}\ }\textbf {\bibinfo {volume} {89}},\
  \bibinfo {pages} {061801} (\bibinfo {year} {2014})}\BibitemShut {NoStop}%
\bibitem [{\citenamefont {Castellanos}\ and\ \citenamefont
  {Escamilla-Rivera}(2017)}]{Castellanos2017}%
  \BibitemOpen
  \bibfield  {author} {\bibinfo {author} {\bibfnamefont {E.}~\bibnamefont
  {Castellanos}}\ and\ \bibinfo {author} {\bibfnamefont {C.}~\bibnamefont
  {Escamilla-Rivera}},\ }\href {\doibase 10.1142/S0217732317500079} {\bibfield
  {journal} {\bibinfo  {journal} {Mod. Phys. Lett. A}\ }\textbf {\bibinfo
  {volume} {32}},\ \bibinfo {pages} {1750007} (\bibinfo {year}
  {2017})}\BibitemShut {NoStop}%
\bibitem [{\citenamefont {Braidotti}\ \emph {et~al.}(2017)\citenamefont
  {Braidotti}, \citenamefont {Musslimani},\ and\ \citenamefont
  {Conti}}]{Braidotti201734}%
  \BibitemOpen
  \bibfield  {author} {\bibinfo {author} {\bibfnamefont {M.~C.}\ \bibnamefont
  {Braidotti}}, \bibinfo {author} {\bibfnamefont {Z.~H.}\ \bibnamefont
  {Musslimani}}, \ and\ \bibinfo {author} {\bibfnamefont {C.}~\bibnamefont
  {Conti}},\ }\href {\doibase http://dx.doi.org/10.1016/j.physd.2016.08.001}
  {\bibfield  {journal} {\bibinfo  {journal} {Physica D}\ }\textbf {\bibinfo
  {volume} {338}},\ \bibinfo {pages} {34 } (\bibinfo {year}
  {2017})}\BibitemShut {NoStop}%
\bibitem [{\citenamefont {{Braidotti}}\ \emph {et~al.}(2018)\citenamefont
  {{Braidotti}}, \citenamefont {{Conti}}, \citenamefont {{Faizal}},
  \citenamefont {{Dey}}, \citenamefont {{Alasfar}}, \citenamefont {{Alsaleh}},\
  and\ \citenamefont {{Ashour}}}]{Faizal2018}%
  \BibitemOpen
  \bibfield  {author} {\bibinfo {author} {\bibfnamefont {M.~C.}\ \bibnamefont
  {{Braidotti}}}, \bibinfo {author} {\bibfnamefont {C.}~\bibnamefont
  {{Conti}}}, \bibinfo {author} {\bibfnamefont {M.}~\bibnamefont {{Faizal}}},
  \bibinfo {author} {\bibfnamefont {S.}~\bibnamefont {{Dey}}}, \bibinfo
  {author} {\bibfnamefont {L.}~\bibnamefont {{Alasfar}}}, \bibinfo {author}
  {\bibfnamefont {S.}~\bibnamefont {{Alsaleh}}}, \ and\ \bibinfo {author}
  {\bibfnamefont {A.}~\bibnamefont {{Ashour}}},\ }\href@noop {} {\bibfield
  {journal} {\bibinfo  {journal} {ArXiv e-prints}\ } (\bibinfo {year}
  {2018})},\ \Eprint {http://arxiv.org/abs/1803.10218} {arXiv:1803.10218}
  \BibitemShut {NoStop}%
\bibitem [{\citenamefont {Carusotto}\ and\ \citenamefont
  {Ciuti}(2013)}]{Carusotto2013}%
  \BibitemOpen
  \bibfield  {author} {\bibinfo {author} {\bibfnamefont {I.}~\bibnamefont
  {Carusotto}}\ and\ \bibinfo {author} {\bibfnamefont {C.}~\bibnamefont
  {Ciuti}},\ }\href {\doibase 10.1103/RevModPhys.85.299} {\bibfield  {journal}
  {\bibinfo  {journal} {Rev. Mod. Phys.}\ }\textbf {\bibinfo {volume} {85}},\
  \bibinfo {pages} {299–366} (\bibinfo {year} {2013})}\BibitemShut {NoStop}%
\bibitem [{\citenamefont {Klaers}\ \emph
  {et~al.}(2010{\natexlab{a}})\citenamefont {Klaers}, \citenamefont
  {Vewinger},\ and\ \citenamefont {Weitz}}]{Klaers2010}%
  \BibitemOpen
  \bibfield  {author} {\bibinfo {author} {\bibfnamefont {J.}~\bibnamefont
  {Klaers}}, \bibinfo {author} {\bibfnamefont {F.}~\bibnamefont {Vewinger}}, \
  and\ \bibinfo {author} {\bibfnamefont {M.}~\bibnamefont {Weitz}},\ }\href
  {http://dx.doi.org/10.1038/nphys1680} {\bibfield  {journal} {\bibinfo
  {journal} {Nat. Phys.}\ }\textbf {\bibinfo {volume} {6}},\ \bibinfo {pages}
  {512} (\bibinfo {year} {2010}{\natexlab{a}})}\BibitemShut {NoStop}%
\bibitem [{\citenamefont {Calvanese~Strinati}\ and\ \citenamefont
  {Conti}(2014)}]{PhysRevA.90.043853}%
  \BibitemOpen
  \bibfield  {author} {\bibinfo {author} {\bibfnamefont {M.}~\bibnamefont
  {Calvanese~Strinati}}\ and\ \bibinfo {author} {\bibfnamefont
  {C.}~\bibnamefont {Conti}},\ }\href {\doibase 10.1103/PhysRevA.90.043853}
  {\bibfield  {journal} {\bibinfo  {journal} {Phys. Rev. A}\ }\textbf {\bibinfo
  {volume} {90}},\ \bibinfo {pages} {043853} (\bibinfo {year}
  {2014})}\BibitemShut {NoStop}%
\bibitem [{\citenamefont {Dominici}\ \emph {et~al.}(2015)\citenamefont
  {Dominici}, \citenamefont {Petrov}, \citenamefont {Matuszewski},
  \citenamefont {Ballarini}, \citenamefont {De~Giorgi}, \citenamefont {Colas},
  \citenamefont {Cancellieri}, \citenamefont {Silva~Fernández}, \citenamefont
  {Bramati}, \citenamefont {Gigli}, \citenamefont {Kavokin}, \citenamefont
  {Laussy},\ and\ \citenamefont {Sanvitto}}]{Dominici2015}%
  \BibitemOpen
  \bibfield  {author} {\bibinfo {author} {\bibfnamefont {L.}~\bibnamefont
  {Dominici}}, \bibinfo {author} {\bibfnamefont {M.}~\bibnamefont {Petrov}},
  \bibinfo {author} {\bibfnamefont {M.}~\bibnamefont {Matuszewski}}, \bibinfo
  {author} {\bibfnamefont {D.}~\bibnamefont {Ballarini}}, \bibinfo {author}
  {\bibfnamefont {M.}~\bibnamefont {De~Giorgi}}, \bibinfo {author}
  {\bibfnamefont {D.}~\bibnamefont {Colas}}, \bibinfo {author} {\bibfnamefont
  {E.}~\bibnamefont {Cancellieri}}, \bibinfo {author} {\bibfnamefont
  {B.}~\bibnamefont {Silva~Fernández}}, \bibinfo {author} {\bibfnamefont
  {A.}~\bibnamefont {Bramati}}, \bibinfo {author} {\bibfnamefont
  {G.}~\bibnamefont {Gigli}}, \bibinfo {author} {\bibfnamefont
  {A.}~\bibnamefont {Kavokin}}, \bibinfo {author} {\bibfnamefont
  {F.}~\bibnamefont {Laussy}}, \ and\ \bibinfo {author} {\bibfnamefont
  {D.}~\bibnamefont {Sanvitto}},\ }\href {http://dx.doi.org/10.1038/ncomms9993}
  {\bibfield  {journal} {\bibinfo  {journal} {Nat. Commun.}\ }\textbf {\bibinfo
  {volume} {6}},\ \bibinfo {pages} {8993} (\bibinfo {year} {2015})}\BibitemShut
  {NoStop}%
\bibitem [{\citenamefont {Amo}\ \emph {et~al.}(2009)\citenamefont {Amo},
  \citenamefont {Lefrère}, \citenamefont {Pigeon}, \citenamefont {Adrados},
  \citenamefont {Ciuti}, \citenamefont {Carusotto}, \citenamefont {Houdré},
  \citenamefont {Giacobino},\ and\ \citenamefont {Bramati}}]{Amo2009}%
  \BibitemOpen
  \bibfield  {author} {\bibinfo {author} {\bibfnamefont {A.}~\bibnamefont
  {Amo}}, \bibinfo {author} {\bibfnamefont {J.}~\bibnamefont {Lefrère}},
  \bibinfo {author} {\bibfnamefont {S.}~\bibnamefont {Pigeon}}, \bibinfo
  {author} {\bibfnamefont {C.}~\bibnamefont {Adrados}}, \bibinfo {author}
  {\bibfnamefont {C.}~\bibnamefont {Ciuti}}, \bibinfo {author} {\bibfnamefont
  {I.}~\bibnamefont {Carusotto}}, \bibinfo {author} {\bibfnamefont
  {R.}~\bibnamefont {Houdré}}, \bibinfo {author} {\bibfnamefont
  {E.}~\bibnamefont {Giacobino}}, \ and\ \bibinfo {author} {\bibfnamefont
  {A.}~\bibnamefont {Bramati}},\ }\href {http://dx.doi.org/10.1038/nphys1364}
  {\bibfield  {journal} {\bibinfo  {journal} {Nat. Phys.}\ }\textbf {\bibinfo
  {volume} {5}},\ \bibinfo {pages} {805} (\bibinfo {year} {2009})}\BibitemShut
  {NoStop}%
\bibitem [{\citenamefont {Chin}\ \emph {et~al.}(2010)\citenamefont {Chin},
  \citenamefont {Grimm}, \citenamefont {Julienne},\ and\ \citenamefont
  {Tiesinga}}]{RevModPhys.82.1225}%
  \BibitemOpen
  \bibfield  {author} {\bibinfo {author} {\bibfnamefont {C.}~\bibnamefont
  {Chin}}, \bibinfo {author} {\bibfnamefont {R.}~\bibnamefont {Grimm}},
  \bibinfo {author} {\bibfnamefont {P.}~\bibnamefont {Julienne}}, \ and\
  \bibinfo {author} {\bibfnamefont {E.}~\bibnamefont {Tiesinga}},\ }\href
  {\doibase 10.1103/RevModPhys.82.1225} {\bibfield  {journal} {\bibinfo
  {journal} {Rev. Mod. Phys.}\ }\textbf {\bibinfo {volume} {82}},\ \bibinfo
  {pages} {1225} (\bibinfo {year} {2010})}\BibitemShut {NoStop}%
\bibitem [{\citenamefont {Longhi}(2018)}]{Longhi2018}%
  \BibitemOpen
  \bibfield  {author} {\bibinfo {author} {\bibfnamefont {S.}~\bibnamefont
  {Longhi}},\ }\href {\doibase 10.1364/OL.43.000226} {\bibfield  {journal}
  {\bibinfo  {journal} {Opt. Lett.}\ }\textbf {\bibinfo {volume} {43}},\
  \bibinfo {pages} {226} (\bibinfo {year} {2018})}\BibitemShut {NoStop}%
\bibitem [{\citenamefont {Das}\ and\ \citenamefont {Vagenas}(2009)}]{Das09}%
  \BibitemOpen
  \bibfield  {author} {\bibinfo {author} {\bibfnamefont {S.}~\bibnamefont
  {Das}}\ and\ \bibinfo {author} {\bibfnamefont {E.~C.}\ \bibnamefont
  {Vagenas}},\ }\href {\doibase 10.1139/P08-105} {\bibfield  {journal}
  {\bibinfo  {journal} {Can.J.Phys.}\ }\textbf {\bibinfo {volume} {87}},\
  \bibinfo {pages} {233–240} (\bibinfo {year} {2009})}\BibitemShut {NoStop}%
\bibitem [{\citenamefont {Briscese}\ \emph {et~al.}(2012)\citenamefont
  {Briscese}, \citenamefont {Grether},\ and\ \citenamefont
  {de~Llano}}]{Briscese12}%
  \BibitemOpen
  \bibfield  {author} {\bibinfo {author} {\bibfnamefont {F.}~\bibnamefont
  {Briscese}}, \bibinfo {author} {\bibfnamefont {M.}~\bibnamefont {Grether}}, \
  and\ \bibinfo {author} {\bibfnamefont {M.}~\bibnamefont {de~Llano}},\ }\href
  {http://stacks.iop.org/0295-5075/98/i=6/a=60001} {\bibfield  {journal}
  {\bibinfo  {journal} {Europhys.Lett.}\ }\textbf {\bibinfo {volume} {98}},\
  \bibinfo {pages} {60001} (\bibinfo {year} {2012})}\BibitemShut {NoStop}%
\bibitem [{\citenamefont {Mercati}\ \emph {et~al.}(2010)\citenamefont
  {Mercati}, \citenamefont {Mazón}, \citenamefont {Amelino-Camelia},
  \citenamefont {Carmona}, \citenamefont {Cortés}, \citenamefont {Induráin},
  \citenamefont {Lämmerzahl},\ and\ \citenamefont {Tino}}]{Mercati10}%
  \BibitemOpen
  \bibfield  {author} {\bibinfo {author} {\bibfnamefont {F.}~\bibnamefont
  {Mercati}}, \bibinfo {author} {\bibfnamefont {D.}~\bibnamefont {Mazón}},
  \bibinfo {author} {\bibfnamefont {G.}~\bibnamefont {Amelino-Camelia}},
  \bibinfo {author} {\bibfnamefont {J.~M.}\ \bibnamefont {Carmona}}, \bibinfo
  {author} {\bibfnamefont {J.~L.}\ \bibnamefont {Cortés}}, \bibinfo {author}
  {\bibfnamefont {J.}~\bibnamefont {Induráin}}, \bibinfo {author}
  {\bibfnamefont {C.}~\bibnamefont {Lämmerzahl}}, \ and\ \bibinfo {author}
  {\bibfnamefont {G.~M.}\ \bibnamefont {Tino}},\ }\href
  {http://stacks.iop.org/0264-9381/27/i=21/a=215003} {\bibfield  {journal}
  {\bibinfo  {journal} {Classical Quantum Grav.}\ }\textbf {\bibinfo {volume}
  {27}},\ \bibinfo {pages} {215003} (\bibinfo {year} {2010})}\BibitemShut
  {NoStop}%
\bibitem [{\citenamefont {Benczik}\ \emph {et~al.}(2002)\citenamefont
  {Benczik}, \citenamefont {Chang}, \citenamefont {Minic}, \citenamefont
  {Okamura}, \citenamefont {Rayyan},\ and\ \citenamefont
  {Takeuchi}}]{PhysRevD.66.026003}%
  \BibitemOpen
  \bibfield  {author} {\bibinfo {author} {\bibfnamefont {S.}~\bibnamefont
  {Benczik}}, \bibinfo {author} {\bibfnamefont {L.~N.}\ \bibnamefont {Chang}},
  \bibinfo {author} {\bibfnamefont {D.}~\bibnamefont {Minic}}, \bibinfo
  {author} {\bibfnamefont {N.}~\bibnamefont {Okamura}}, \bibinfo {author}
  {\bibfnamefont {S.}~\bibnamefont {Rayyan}}, \ and\ \bibinfo {author}
  {\bibfnamefont {T.}~\bibnamefont {Takeuchi}},\ }\href {\doibase
  10.1103/PhysRevD.66.026003} {\bibfield  {journal} {\bibinfo  {journal} {Phys.
  Rev. D}\ }\textbf {\bibinfo {volume} {66}},\ \bibinfo {pages} {026003}
  (\bibinfo {year} {2002})}\BibitemShut {NoStop}%
\bibitem [{\citenamefont {Brau}(1999)}]{Brau1999}%
  \BibitemOpen
  \bibfield  {author} {\bibinfo {author} {\bibfnamefont {F.}~\bibnamefont
  {Brau}},\ }\href {http://stacks.iop.org/0305-4470/32/i=44/a=308} {\bibfield
  {journal} {\bibinfo  {journal} {J. Phys. A: Math. Gen.}\ }\textbf {\bibinfo
  {volume} {32}},\ \bibinfo {pages} {7691} (\bibinfo {year}
  {1999})}\BibitemShut {NoStop}%
\bibitem [{\citenamefont {Brown}(1972)}]{Brown1972}%
  \BibitemOpen
  \bibfield  {author} {\bibinfo {author} {\bibfnamefont {L.~S.}\ \bibnamefont
  {Brown}},\ }\href {\doibase 10.1119/1.1986554} {\bibfield  {journal}
  {\bibinfo  {journal} {Am. J. Phys.}\ }\textbf {\bibinfo {volume} {40}},\
  \bibinfo {pages} {371} (\bibinfo {year} {1972})}\BibitemShut {NoStop}%
\bibitem [{\citenamefont {Van~Vleck}(1928)}]{VanVleck1928}%
  \BibitemOpen
  \bibfield  {author} {\bibinfo {author} {\bibfnamefont {J.~H.}\ \bibnamefont
  {Van~Vleck}},\ }\href {\doibase 10.1073/pnas.14.2.178} {\bibfield  {journal}
  {\bibinfo  {journal} {Proc. N. A. S.}\ }\textbf {\bibinfo {volume} {14}},\
  \bibinfo {pages} {178} (\bibinfo {year} {1928})}\BibitemShut {NoStop}%
\bibitem [{\citenamefont {Dalfovo}\ \emph {et~al.}(1999)\citenamefont
  {Dalfovo}, \citenamefont {Giorgini}, \citenamefont {Pitaevskii},\ and\
  \citenamefont {Stringari}}]{Dalfovo99}%
  \BibitemOpen
  \bibfield  {author} {\bibinfo {author} {\bibfnamefont {F.}~\bibnamefont
  {Dalfovo}}, \bibinfo {author} {\bibfnamefont {S.}~\bibnamefont {Giorgini}},
  \bibinfo {author} {\bibfnamefont {L.}~\bibnamefont {Pitaevskii}}, \ and\
  \bibinfo {author} {\bibfnamefont {S.}~\bibnamefont {Stringari}},\ }\href@noop
  {} {\bibfield  {journal} {\bibinfo  {journal} {Rev. Mod. Phys.}\ }\textbf
  {\bibinfo {volume} {71}},\ \bibinfo {pages} {463–512} (\bibinfo {year}
  {1999})}\BibitemShut {NoStop}%
\bibitem [{\citenamefont {Landau}\ and\ \citenamefont
  {Lifshitz}(2013)}]{landau2013QuaMecShoCouThePhy}%
  \BibitemOpen
  \bibfield  {author} {\bibinfo {author} {\bibfnamefont {L.~D.}\ \bibnamefont
  {Landau}}\ and\ \bibinfo {author} {\bibfnamefont {E.~M.}\ \bibnamefont
  {Lifshitz}},\ }\href
  {http://books.google.it/books?hl=en&lr=&id=oOFbAwAAQBAJ&oi=fnd&pg=PP1&dq=quantum+mechanics+landau&ots=hb7XXoefze&sig=92SBOJi6U-679wtP8KwI28ySgu4}
  {\emph {\bibinfo {title} {{Quantum Mechanics: A Shorter Course of Theoretical
  Physics}}}}\ (\bibinfo  {publisher} {Elsevier},\ \bibinfo {year}
  {2013})\BibitemShut {NoStop}%
\bibitem [{\citenamefont {Goldstein}(1965)}]{goldsteinbook}%
  \BibitemOpen
  \bibfield  {author} {\bibinfo {author} {\bibfnamefont {H.}~\bibnamefont
  {Goldstein}},\ }\href@noop {} {\emph {\bibinfo {title} {Classical
  Mechanics}}}\ (\bibinfo  {publisher} {Pearson Education India},\ \bibinfo
  {year} {1965})\BibitemShut {NoStop}%
\bibitem [{\citenamefont {Rauschenbakh}\ \emph {et~al.}(2002)\citenamefont
  {Rauschenbakh}, \citenamefont {M.~Y.~Ovchinnikov},\ and\ \citenamefont
  {McKenna-Lawlor}}]{bookspaceflight}%
  \BibitemOpen
  \bibfield  {author} {\bibinfo {author} {\bibfnamefont {V.}~\bibnamefont
  {Rauschenbakh}}, \bibinfo {author} {\bibfnamefont {M.~Y.}\ \bibnamefont
  {M.~Y.~Ovchinnikov}}, \ and\ \bibinfo {author} {\bibfnamefont {S.~M.~P.}\
  \bibnamefont {McKenna-Lawlor}},\ }\href@noop {} {\emph {\bibinfo {title}
  {Essential Spaceflight Dynamics and Magnetospherics}}}\ (\bibinfo
  {publisher} {Springer},\ \bibinfo {year} {2002})\BibitemShut {NoStop}%
\bibitem [{\citenamefont {Einstein}(1915)}]{Einstein1915}%
  \BibitemOpen
  \bibfield  {author} {\bibinfo {author} {\bibfnamefont {A.}~\bibnamefont
  {Einstein}},\ }\href@noop {} {\emph {\bibinfo {title} {Erklarung der
  Perihelwegung Merkur der allgemeinen Relativitatstheorie}}}\ (\bibinfo
  {publisher} {Koniglich Preussische Akademie der Wissenschaften},\ \bibinfo
  {year} {1915})\BibitemShut {NoStop}%
\bibitem [{\citenamefont {Kox}\ and\ \citenamefont
  {Schutman}(1996)}]{CollectedEinstein}%
  \BibitemOpen
  \bibinfo {editor} {\bibfnamefont {J.}~\bibnamefont {Kox}}\ and\ \bibinfo
  {editor} {\bibfnamefont {R.}~\bibnamefont {Schutman}},\ eds.,\ \href@noop {}
  {\emph {\bibinfo {title} {The collected papers of Albert Einstein}}}\
  (\bibinfo  {publisher} {Princeton University Press},\ \bibinfo {year}
  {1996})\BibitemShut {NoStop}%
\bibitem [{\citenamefont {Wisner}\ \emph {et~al.}(1973)\citenamefont {Wisner},
  \citenamefont {Thorne},\ and\ \citenamefont {Wheeler}}]{MTWbook}%
  \BibitemOpen
  \bibfield  {author} {\bibinfo {author} {\bibfnamefont {C.~W.}\ \bibnamefont
  {Wisner}}, \bibinfo {author} {\bibfnamefont {K.~S.}\ \bibnamefont {Thorne}},
  \ and\ \bibinfo {author} {\bibfnamefont {J.~A.}\ \bibnamefont {Wheeler}},\
  }\href@noop {} {\emph {\bibinfo {title} {Gravitation}}}\ (\bibinfo
  {publisher} {W. H. Freeman},\ \bibinfo {year} {1973})\BibitemShut {NoStop}%
\bibitem [{\citenamefont {Yao}\ and\ \citenamefont
  {Padgett}(2011)}]{Padgett2011}%
  \BibitemOpen
  \bibfield  {author} {\bibinfo {author} {\bibfnamefont {A.~M.}\ \bibnamefont
  {Yao}}\ and\ \bibinfo {author} {\bibfnamefont {M.~J.}\ \bibnamefont
  {Padgett}},\ }\href@noop {} {\bibfield  {journal} {\bibinfo  {journal} {Adv.
  Opt. Photonics}\ }\textbf {\bibinfo {volume} {3}},\ \bibinfo {pages} {161}
  (\bibinfo {year} {2011})}\BibitemShut {NoStop}%
\bibitem [{\citenamefont {Yariv}(1991)}]{YarivBook}%
  \BibitemOpen
  \bibfield  {author} {\bibinfo {author} {\bibfnamefont {A.}~\bibnamefont
  {Yariv}},\ }\href@noop {} {\emph {\bibinfo {title} {{Quantum Electronics}}}}\
  (\bibinfo  {publisher} {Saunders College, San Diego},\ \bibinfo {year}
  {1991})\BibitemShut {NoStop}%
\bibitem [{\citenamefont {Helmerson}\ \emph {et~al.}(2007)\citenamefont
  {Helmerson}, \citenamefont {Andersen}, \citenamefont {Ryu}, \citenamefont
  {Cladé}, \citenamefont {Natarajan}, \citenamefont {Vaziri},\ and\
  \citenamefont {Phillips}}]{Helmerson2007}%
  \BibitemOpen
  \bibfield  {author} {\bibinfo {author} {\bibfnamefont {K.}~\bibnamefont
  {Helmerson}}, \bibinfo {author} {\bibfnamefont {M.}~\bibnamefont {Andersen}},
  \bibinfo {author} {\bibfnamefont {C.}~\bibnamefont {Ryu}}, \bibinfo {author}
  {\bibfnamefont {P.}~\bibnamefont {Cladé}}, \bibinfo {author} {\bibfnamefont
  {V.}~\bibnamefont {Natarajan}}, \bibinfo {author} {\bibfnamefont
  {A.}~\bibnamefont {Vaziri}}, \ and\ \bibinfo {author} {\bibfnamefont
  {W.}~\bibnamefont {Phillips}},\ }\href {\doibase
  https://doi.org/10.1016/j.nuclphysa.2007.03.019} {\bibfield  {journal}
  {\bibinfo  {journal} {Nuclear Physics A}\ }\textbf {\bibinfo {volume}
  {790}},\ \bibinfo {pages} {705c } (\bibinfo {year} {2007})},\ \bibinfo {note}
  {few-Body Problems in Physics}\BibitemShut {NoStop}%
\bibitem [{\citenamefont {Dall}\ \emph {et~al.}(2014)\citenamefont {Dall},
  \citenamefont {Fraser}, \citenamefont {Desyatnikov}, \citenamefont {Li},
  \citenamefont {Brodbeck}, \citenamefont {Kamp}, \citenamefont {Schneider},
  \citenamefont {H\"ofling},\ and\ \citenamefont {Ostrovskaya}}]{Dall2014}%
  \BibitemOpen
  \bibfield  {author} {\bibinfo {author} {\bibfnamefont {R.}~\bibnamefont
  {Dall}}, \bibinfo {author} {\bibfnamefont {M.~D.}\ \bibnamefont {Fraser}},
  \bibinfo {author} {\bibfnamefont {A.~S.}\ \bibnamefont {Desyatnikov}},
  \bibinfo {author} {\bibfnamefont {G.}~\bibnamefont {Li}}, \bibinfo {author}
  {\bibfnamefont {S.}~\bibnamefont {Brodbeck}}, \bibinfo {author}
  {\bibfnamefont {M.}~\bibnamefont {Kamp}}, \bibinfo {author} {\bibfnamefont
  {C.}~\bibnamefont {Schneider}}, \bibinfo {author} {\bibfnamefont
  {S.}~\bibnamefont {H\"ofling}}, \ and\ \bibinfo {author} {\bibfnamefont
  {E.~A.}\ \bibnamefont {Ostrovskaya}},\ }\href {\doibase
  10.1103/PhysRevLett.113.200404} {\bibfield  {journal} {\bibinfo  {journal}
  {Phys. Rev. Lett.}\ }\textbf {\bibinfo {volume} {113}},\ \bibinfo {pages}
  {200404} (\bibinfo {year} {2014})}\BibitemShut {NoStop}%
\bibitem [{\citenamefont {Boulier}\ \emph {et~al.}(2016)\citenamefont
  {Boulier}, \citenamefont {Cancellieri}, \citenamefont {Sangouard},
  \citenamefont {Glorieux}, \citenamefont {Kavokin}, \citenamefont {Whittaker},
  \citenamefont {Giacobino},\ and\ \citenamefont {Bramati}}]{Bramati2016}%
  \BibitemOpen
  \bibfield  {author} {\bibinfo {author} {\bibfnamefont {T.}~\bibnamefont
  {Boulier}}, \bibinfo {author} {\bibfnamefont {E.}~\bibnamefont
  {Cancellieri}}, \bibinfo {author} {\bibfnamefont {N.~D.}\ \bibnamefont
  {Sangouard}}, \bibinfo {author} {\bibfnamefont {Q.}~\bibnamefont {Glorieux}},
  \bibinfo {author} {\bibfnamefont {A.~V.}\ \bibnamefont {Kavokin}}, \bibinfo
  {author} {\bibfnamefont {D.~M.}\ \bibnamefont {Whittaker}}, \bibinfo {author}
  {\bibfnamefont {E.}~\bibnamefont {Giacobino}}, \ and\ \bibinfo {author}
  {\bibfnamefont {A.}~\bibnamefont {Bramati}},\ }\href {\doibase
  10.1103/PhysRevLett.116.116402} {\bibfield  {journal} {\bibinfo  {journal}
  {Phys. Rev. Lett.}\ }\textbf {\bibinfo {volume} {116}},\ \bibinfo {pages}
  {116402} (\bibinfo {year} {2016})}\BibitemShut {NoStop}%
\bibitem [{\citenamefont {Klaers}\ \emph
  {et~al.}(2010{\natexlab{b}})\citenamefont {Klaers}, \citenamefont {Schmitt},
  \citenamefont {Vewinger},\ and\ \citenamefont {Weitz}}]{weitz2010}%
  \BibitemOpen
  \bibfield  {author} {\bibinfo {author} {\bibfnamefont {J.}~\bibnamefont
  {Klaers}}, \bibinfo {author} {\bibfnamefont {J.}~\bibnamefont {Schmitt}},
  \bibinfo {author} {\bibfnamefont {F.}~\bibnamefont {Vewinger}}, \ and\
  \bibinfo {author} {\bibfnamefont {M.}~\bibnamefont {Weitz}},\ }\href
  {\doibase 10.1038/nature09567} {\bibfield  {journal} {\bibinfo  {journal}
  {Nature}\ }\textbf {\bibinfo {volume} {468}},\ \bibinfo {pages} {545}
  (\bibinfo {year} {2010}{\natexlab{b}})}\BibitemShut {NoStop}%
\bibitem [{NAS(2018)}]{NASACAL}%
  \BibitemOpen
  \href@noop {} {}\bibinfo {howpublished}
  {\url{https://coldatomlab.jpl.nasa.gov}} (\bibinfo {year} {2018})\BibitemShut
  {NoStop}%
\end{thebibliography}
%

\end{document}